\documentclass[a4paper,11pt]{article}
\usepackage[dvipsnames]{xcolor}% load first: jinstpub/pgfplots pull in xcolor with no options, so loading it later causes an option clash
\usepackage{jinstpub}
\usepackage{graphicx}
\usepackage{dcolumn}
\usepackage{bm}
\usepackage{subfigure,dcolumn}
\usepackage{hyperref}
\usepackage{pgfplots}
\usepackage{booktabs}
\usepackage{multirow}
\pgfplotsset{compat=1.18}

\title{\boldmath Robust Betatron-Tune Measurement from Schottky Spectra: Complementary Classical and Deep-Learning Paradigms}

\author[a,b]{P. Sun}
\author[a,c,1]{M. Zhang\note{Corresponding author.}}
\author[a,c]{R. Yuan}
\author[a]{D. Li}
\author[c]{and J. Dong}

\affiliation[a]{Shanghai Institute of Applied Physics, Chinese Academy of Sciences,\\
Shanghai 201800, China}
\affiliation[b]{University of Chinese Academy of Sciences,\\
Beijing 100049, China}
\affiliation[c]{Shanghai Advanced Research Institute, Chinese Academy of Sciences,\\
Shanghai 201204, China}

\emailAdd{zhangmanzhou@sinap.ac.cn}

\abstract{
Schottky spectra carry rich diagnostic information about circulating beams, including chromaticity and emittance, with the betatron sidebands encoding the fractional tune. Reliable tune readout is essential for third-order resonance slow extraction in compact medical proton synchrotrons. In this setting, hardware constraints produce low signal-to-noise ratios and limited frequency resolution, so classical algorithms such as peak detection and curve fitting are readily diverted. This work addresses this challenge with two complementary estimators that share a common spectral front-end but differ in how they carry temporal context. The classical estimator pools spectral evidence across frames through a motion-compensated coherent moving average that aggregates the moving sideband, localises the betatron sideband with a multi-width matched-filter bank, and resolves the tune at sub-bin precision with a local-window argmax and an adaptive-window, median-absolute-deviation-gated centroid sub-bin readout on the pooled spectrum. The deep-learning estimator maps each preprocessed spectrum to a tune-grid likelihood map via a convolutional neural network with global convolutions evaluated by the fast Fourier transform, propagates the posterior with a discrete two-dimensional $(q, v)$ Bayes tracker under a Gaussian motion model, taking the network likelihood map as its per-frame observation, and reports a per-frame posterior standard deviation. The two estimators thus accumulate temporal context through structurally distinct representations. Across a synthetic dynamic-tune benchmark, the deep-learning estimator substantially outperforms both published baselines while the classical estimator surpasses the latency-compensated baseline; the two are complementary on deployment rather than accuracy grounds, with the deep-learning estimator the more accurate across the operating envelope, while the classical estimator runs on a single CPU core with no training data. On near-stationary SAPT real-beam data, both estimators run end-to-end (the deep-learning estimator without retraining), and median per-frame latency remains sub-millisecond on commodity hardware, indicating the potential for real-time-capable operation at compact medical synchrotrons such as SAPT.

}


\arxivnumber{TBD}

\begin{document}
\maketitle
\flushbottom

\section{Introduction}
\label{sec:intro}
Compact medical proton synchrotrons that rely on third-order resonance slow extraction impose stringent requirements on betatron-tune measurement. Drifts in the fractional tune $q$ directly affect spill quality, and the measurement must track energy ramps that compress the per-frame acquisition window to a millisecond scale. At the Shanghai Advanced Proton Therapy (SAPT) facility~\cite{zhang2023sapt}, the limited pickup-electrode length, beam-position-monitor (BPM) sensitivity, and beam current together reduce the Schottky sideband power available for tune extraction. This places SAPT below the signal-quality regime in which Large Hadron Collider Schottky workflows are normally reported~\cite{lasochaonline,lasocha2024jacow}. The short acquisition windows imposed by ramping and extraction further limit the frequency resolution, and narrow-band lines from clocking, switching, or RF pickup may remain in the measured spectra when they are not fully suppressed at the analogue front end.

At present, SAPT obtains the tune by exciting the beam with the slow-extraction kicker and applying a Fourier transform to the resulting BPM signal. In practice, the residual post-injection oscillations decay too rapidly to yield a stable coherent tune line, and the excitation disturbs the slow-extraction process. Schottky-based diagnostics~\cite{van1972stochastic,van2005diagnostics,caspers2009schottky} offer a feasible alternative by passively recording the spectral imprint of incoherent transverse motion. On compact medical machines, however, classical sideband methods reported in the literature, including peak detection for the coherent tune, curve fitting for the incoherent tune, and a mirrored-difference method exploiting the residual power correlation between corresponding positive and negative Bessel satellites around a transverse betatron sideband~\cite{betz2017bunched,lasocha2022estimation}, typically assume more favourable signal quality and longer acquisition or averaging windows than are available here.

Our two prior works~\cite{sun2025high,sun2026real} together established the upstream preprocessing pipeline used here, with~\cite{sun2026real} building on~\cite{sun2025high}. They further demonstrate that, under SAPT-relevant conditions such as low signal-to-noise ratios (SNR) and short acquisition windows, frame-level tune extraction is achievable with both classical signal processing and a lightweight learned estimator when each acts on its own preprocessed input. This work therefore prioritises real-time latency and accuracy under realistic noise conditions.

We now isolate the downstream tune estimator. With the preprocessing chain held fixed, the remaining tune-estimation error is determined by the per-frame estimator that consumes the preprocessed spectrum and emits a tune. On this common input, stateless peak detection, curve fitting, and weighted-centroid baselines can be diverted in the SAPT-relevant operating regimes considered here, motivating estimators that accumulate temporal evidence without losing responsiveness to tune motion.

This paper investigates two complementary tune estimators that share a common spectral front-end (the upstream preprocessing pipeline) but address the temporal-context problem through structurally different representations. The classical estimator carries temporal memory entirely in the power-spectrum domain, whereas the deep-learning estimator carries it in a discrete probabilistic tune-velocity state space. The two estimators are complementary on deployment rather than accuracy grounds: the deep-learning estimator is the more accurate across the operating envelope, while the classical estimator is training-free and runs on a single CPU core. The specific contributions are:
\begin{enumerate}
    \item \textbf{Classical tune estimator.} A motion-compensated coherent moving average with a motion-adaptive pool depth aggregates spectral evidence across frames. A multi-width matched-filter bank locates the betatron sideband, with a local-window argmax selecting the integer peak and an adaptive-window median absolute deviation (MAD)-gated centroid on the pooled spectrum resolving it to sub-bin precision. The pipeline reaches a median per-frame latency of approximately $0.12$\,ms on a single CPU core (Table~\ref{tab:latency}).
    \item \textbf{Deep-learning tune estimator.} A convolutional neural network (CNN) with global convolutions evaluated using the fast Fourier transform (FFT) maps each preprocessed power spectrum to a tune-grid likelihood map. A discrete two-dimensional $(q, v)$ Bayes tracker propagates the temporal posterior under a Gaussian motion model and updates it with the network likelihood map as the per-frame observation, with the marginal tune posterior providing both the per-frame estimate and posterior standard deviation. After graph compilation, the pipeline reaches a median per-frame latency of approximately $0.40$\,ms on a commodity graphics processing unit (GPU) (Table~\ref{tab:latency}).
\end{enumerate}

Section~\ref{sec:results} shows that the two proposed estimators substantially outperform published baselines on a synthetic dynamic-tune benchmark and run end-to-end without retraining on near-stationary SAPT real-beam data. The cell-by-cell pattern of which estimator wins where further motivates an operating-regime decision matrix rather than a universal-winner claim.

The paper is organised as follows. Section~\ref{sec:methods} establishes the notation for the Schottky spectrum and folded tune, describes the shared upstream preprocessing chain, and introduces the two tune estimators in parallel within a minimal common framework to make their interfaces directly comparable. Section~\ref{sec:results} covers a static single-frame benchmark, the main synthetic dynamic-tune benchmark across SNR and trajectory shape, robustness studies spanning beam-loss recovery, residual narrow-band interference, and out-of-distribution SNR, component ablations and per-frame latency, a preliminary validation on SAPT real-beam data, and the operating-regime decision matrix. Section~\ref{sec:conclusions} summarises the findings and outlines limitations and future work.

\section{Background and methods}
\label{sec:methods}
This section fixes the Schottky-spectrum and folded-tune notation, describes the shared upstream preprocessing chain and the interface both estimators consume, and then develops the two estimators in parallel within a minimal common framework.

\subsection{Schottky signals and the folded-tune readout}
\label{subsec:schottky}
\paragraph{Schottky signal model.}
The Schottky spectrum of a circulating beam is the random spectral signature of single-particle motion as picked up by a stripline or button electrode~\cite{boussard1986schottky,chanon2016schottky,nolden2001instrumentation,lasocha2020estimation}. A circulating particle of charge $e$ at revolution frequency $f_{\text{rev}}$ deposits a near-delta pulse train on the electrode at intervals $T = 1/f_{\text{rev}}$; the corresponding Fourier components fall at integer multiples of $f_{\text{rev}}$,
\begin{align}
\label{eq:harmonics}
    f_n = n\, f_{\text{rev}},
\end{align}
indexed by the harmonic number $n$. Transverse betatron motion adds an amplitude modulation that splits each harmonic into the symmetric pair
\begin{align}
\label{eq:sidebands}
    f_{n,\pm q} = (n \pm q)\, f_{\text{rev}},
\end{align}
where $q \in [0, 1)$ is the fractional betatron tune. Because $n+q$ and $n-q$ sit symmetrically about $f_n$, the offset of the closer sideband from the nearest harmonic identifies the folded tune $q_{05} = \min(q, 1 - q)$; we use this folded coordinate on $[0, 0.5)$ as the reported quantity throughout the rest of the paper.

In a coasting beam the principal harmonics and the betatron sidebands appear as continuous distributions whose shape encodes the underlying momentum spread and betatron-amplitude distribution. A bunched beam introduces a further frequency component at the synchrotron frequency $f_s$, which sprays each principal sideband into a series of Bessel satellites,
\begin{align}
\label{eq:bessel}
    f_{n,\pm q,\pm k} = \bigl(n \pm q \pm k\, f_s / f_{\text{rev}}\bigr)\, f_{\text{rev}},
    \qquad k \in \mathbb{Z},
\end{align}
the longitudinal-line structure of which is not used by the tune estimators of this work. In either regime the tune information of interest stays in the offset of the closest sideband from the nearest harmonic, so the network is trained on the folded tune coordinate without distinguishing the two beam modes. After the preprocessing pipeline of Section~\ref{subsec:preprocessing} maps the spectrum onto a uniform $L = 1024$-bin grid spanning $[0, 0.5)$, harmonics collapse near $q = 0$ while transverse sidebands appear at their folded offsets, so tune extraction reduces to a sideband-localisation task on a fixed grid.

\paragraph{Measurement model.}
Each per-frame power spectral density (PSD) $S_t(f)$ in this work is treated as a sum of a Schottky contribution at the operating tune $q_t$ and an additive noise term that bundles broadband electronics noise with any sparse narrow-band lines that may leak in from the front-end chain (digital clocks, power supplies, RF systems),
\begin{align}
\label{eq:psd_obs}
    S_t(f) = S_t^{\text{sig}}(f) + S_t^{\text{noise}}(f).
\end{align}
The SNR convention here matches the companion work~\cite{sun2026real}: time-domain Gaussian noise is calibrated to a prescribed ratio between single-sideband Schottky power and noise power inside half the revolution-frequency bandwidth. Sparse impulsive lines are added at randomised positions and amplitudes within calibrated ranges; these lines act as synthetic surrogates for possible narrow-band interference in operational Schottky chains rather than as features drawn from measured spectra.

\subsection{Upstream preprocessing pipeline}
\label{subsec:preprocessing}
\begin{figure}
    \centering
    \includegraphics[width=0.9\linewidth]{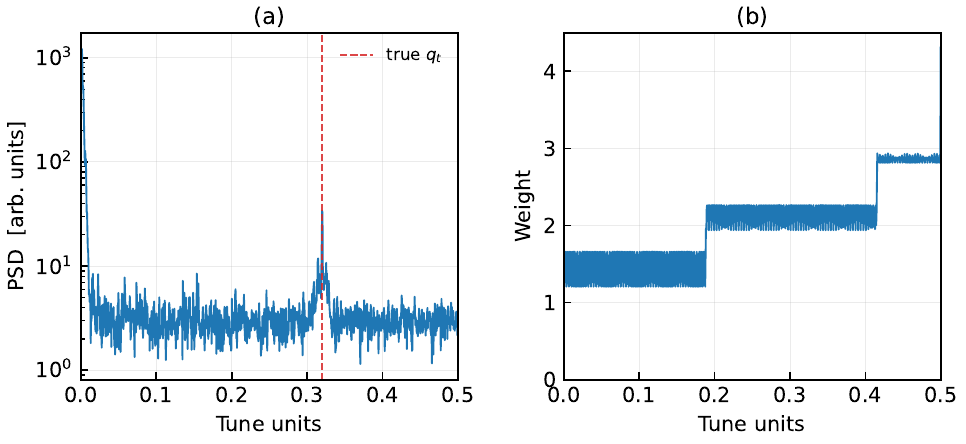}
    \caption{Representative input pair to the downstream estimators after SAPT-side preprocessing, plotted on a coasting-beam frame at $\mathrm{SNR} \approx -10$\,dB, sampling rate $16$\,MHz, and revolution frequency $\approx 5.7$\,MHz. (a)~Folded power spectrum $x[k]$ on the $L = 1024$-bin tune grid (logarithmic vertical scale to span the residual DC content and the betatron sideband on the same axis); the dashed line marks the ground-truth tune $q_t$. (b)~Per-bin sampling-weight map $w[k]$ derived from linear soft binning, encoding the non-uniform sampling density induced by frequency-to-tune mapping. The downstream classical estimator (Section~\ref{subsec:emasc}) consumes $x$ alone; the deep-learning estimator (Section~\ref{subsec:nlbf}) consumes both $x$ and $w$.}
    \label{fig:xw_grid}
\end{figure}

Both estimators consume per-frame measurements that have already been mapped onto the half-open folded-tune grid $q \in [0, 0.5)$ by the SAPT preprocessing pipeline introduced in~\cite{sun2026real}. In summary, the absolute-frequency PSD is expressed relative to adjacent revolution harmonics, folded into the half interval, and accumulated on a uniform $L = 1024$-point grid by linear soft binning. Each frequency component contributes to its two neighbouring grid points using the interpolation weights, producing both a folded PSD $x[k]$ and a per-bin sampling-weight map $w[k]$. The pair $(x, w)$ constitutes the per-frame input to the downstream estimators, with the classical estimator in Section~\ref{subsec:emasc} consuming $x$ alone and the deep-learning estimator in Section~\ref{subsec:nlbf} consuming both $x$ and $w$ via its feature builder. Figure~\ref{fig:xw_grid} illustrates a representative $(x, w)$ pair after preprocessing, where the betatron sideband appears as a localised peak on the folded tune axis.

\subsection{Common per-frame interface}
\label{subsec:common_interface}
The two estimators in Sections~\ref{subsec:emasc} and~\ref{subsec:nlbf} consume the same per-frame input: the post-mapped folded PSD $S_t \in \mathbb{R}^{L}$ on the $L = 1024$-bin tune grid produced by the preprocessing pipeline of Section~\ref{subsec:preprocessing}, together with the optional per-bin sampling-weight map $w_t \in \mathbb{R}^{L}$ and a per-bin low-tune validity mask $m \in \{0, 1\}^{L}$ that excludes the region near the folded longitudinal harmonics; in the deployed pipeline the deep-learning estimator applies this mask as a static $q$-grid cut. Each estimator emits, at every frame, a folded-tune estimate $\hat q_t \in [0, 0.5)$. The classical estimator does not consume the sampling-weight map: its sub-bin centroid is read relative to a local floor set by the window median and median absolute deviation, while the matched-filter detector operates on the pooled-PSD response, leaving the readout insensitive to the per-bin sampling density. The deep-learning estimator does consume the weight map: its trained feature builder treats it as one of three input channels, alongside the normalised PSD and a tune-coordinate channel, and the per-bin sampling density encoded by $w$ modulates the local statistical noise that the network can learn to exploit alongside the spectral structure (see Section~\ref{subsec:nlbf}).

\subsection{Classical estimator: motion-compensated PSD-domain temporal context}
\label{subsec:emasc}
\paragraph{Postprocess block.}\label{para:emasc-postprocess}
Before the temporal-context blocks below consume each new frame, the mapped PSD $x$ is passed through a stateless postprocess block that follows the low-tune-mask plus fixed-width PSD smoother design of~\cite{sun2025high}, specialised here with a Savitzky-Golay polynomial smoother in place of the Gaussian kernel of~\cite{sun2025high}. The block performs two operations in fixed order. First, in the deployed configuration the low-tune mask zeroes bins flagged invalid by the upstream pipeline together with a small region near $q = 0$, suppressing residual DC leakage and longitudinal-harmonic bleed; the mask is part of the postprocess block in deployments where upstream channel decoupling does not fully attenuate the low-tune region. Second, a Savitzky-Golay polynomial filter~\cite{schafer2011savitzky,betz2017bunched} of window length fixed in tune units, $W_{\mathrm{SG}} = 10^{-2}$, converted to the nearest odd bin count on the operating grid ($21$ bins at the deployed $L = 1024$), and polynomial order $p_{\mathrm{SG}} = 2$, a second-order fit that preserves the local curvature of the betatron sideband, suppresses broadband noise while better preserving the local curvature and amplitude of the betatron sideband than a Gaussian smoother of equal width, with polynomial-interpolation boundary handling. The polynomial-fit smoother is preferred over a Gaussian kernel of equivalent width because a Gaussian kernel broadens every local feature toward its own scale, reducing the shape diversity that the downstream multi-width Gaussian matched-filter bank exploits on the smoothed PSD. The mask is reapplied after the smoother so that the polynomial fit cannot leak non-zero amplitude back into bins it had zeroed. The output of this block is the per-frame postprocessed PSD $\tilde S$ that feeds the temporal-context blocks below.

The matched-filter bank, local-window argmax with Shewhart-gated global fallback~\cite{shewhart2022economic}, and adaptive-window MAD-gated centroid sub-bin readout described in this subsection operate on a generic tune-axis PSD input $S^* \in \mathbb{R}^{L}$: the matched-filter bank forms a response on which the integer peak is located, while the sub-bin centroid is evaluated on $S^*$ itself. In the deployed pipeline (the deployed-pipeline ordering described below) the input $S^*$ is the motion-compensated coherent moving average of the per-frame postprocessed PSDs $\tilde S_t$, and we treat it as a fixed input for the purpose of stating the readout.

\paragraph{Matched-filter bank.}
Sideband shapes on the mapped PSD vary with the upstream chain configuration and the harmonic number, with effective widths from one or two bins on narrow lines up to a few tens of bins on strongly broadened sidebands. To localise a peak whose width is not known a priori, a bank of $K = 4$ energy-normalised Gaussian kernels of dyadic widths $\sigma_q \in \{0.5, 1, 2, 4\} \times 10^{-3}$ tune units is convolved into $S^*$, each width converted to its bin-domain value $\sigma = \sigma_q / (0.5/L)$ on the operating grid, so that the bank follows the tune-axis resolution rather than a fixed bin geometry (approximately $\{1, 2, 4, 8\}$ bins at the deployed $L = 1024$). The dyadic spacing covers the range of betatron-sideband widths encountered in practice, from the narrowest near-single-bin lines to the broadened sidebands at the SAPT operating point. At each tune-axis bin, the response is aggregated as the mean of the $\lceil K / 2 \rceil = 2$ largest kernel responses,
\begin{align}
\label{eq:mf_response}
    M[k] \;=\; \frac{1}{\lceil K/2 \rceil}\,\sum_{\sigma \in \mathcal{T}_k}\!\bigl(S^* \ast h_\sigma\bigr)[k],
    \qquad
    h_\sigma[k] \,\propto\, \exp\!\Bigl(-\tfrac{k^2}{2\sigma^2}\Bigr),\quad \lVert h_\sigma\rVert_2 = 1,
\end{align}
where $\mathcal{T}_k$ is the set of $\lceil K/2 \rceil$ kernels with the largest response at bin $k$, each kernel is truncated to $\pm 3\sigma$ samples (a $6\sigma$ full-width window that captures more than $99.7\%$ of the Gaussian mass), and the $L_2$ normalisation $\lVert h_\sigma\rVert_2 = 1$ places the responses to different widths on a common scale. The upper-half-mean is a lower-tail-robust aggregator over the kernel bank: it discards the two weakest width responses and combines the two strongest at each bin, in contrast to a pointwise maximum that locks onto a single kernel and a pointwise mean that washes the response with badly mismatched kernels. Each convolution runs in time linear in $L$ with a small constant factor determined by the kernel width, and the four widths together account for a few tens of microseconds of the per-frame budget.

\paragraph{Local-window argmax.}
The readout selects a window of half-width $W_q$ bins around the previous tune $\hat q_{t-1}$ on the bin-index axis, where $W_q = \mathrm{round}(w_q / \Delta_q)$ converts a half-width fixed in tune units, $w_q = 10^{-2}$, to the operating grid ($W_q = 20$ bins at the deployed $L = 1024$), so that the search envelope, like the matched-filter bank above, follows the tune-axis resolution rather than a fixed bin geometry:
\begin{align}
\label{eq:window}
    \mathcal{W}_t \;=\; \bigl\{\,k\in\{0,\dots,L-1\} \,:\, |k - k_{t-1}| \le W_q\,\bigr\},
    \qquad k_{t-1} = \mathrm{round}\!\bigl(\hat q_{t-1}/\Delta_q\bigr),
\end{align}
where $\Delta_q = 0.5 / L$ is the tune-axis bin width.
The window restricts the readout to a tune-velocity envelope compatible with the trajectories considered in Section~\ref{sec:results}: the half-width $w_q = 10^{-2}$ admits a single-frame deflection of one fiftieth of the folded-tune range at every grid size, well above the per-frame tune motion encountered during steady-state slow extraction and acceleration ramps at compact medical synchrotrons. Inside $\mathcal{W}_t$ the readout takes the matched-filter argmax,
\begin{align}
\label{eq:local_argmax}
    \hat k_t^{\mathrm{loc}} \;=\; \arg\max_{k \in \mathcal{W}_t} M[k],
\end{align}
which yields the integer local peak $\hat k_t^{\mathrm{loc}}$. The selection between this local peak and the global matched-filter peak is made by the runtime-noise-gated fallback below, and a single sub-bin refinement is then applied to the selected integer bin.

\paragraph{Jump-to-global fallback with runtime-noise gating.}
When a sustained tune motion or an external machine event drives the true sideband outside $\mathcal{W}_t$, the windowed argmax above is unable to recover until the prior catches up; we use a runtime estimate of the matched-filter response's own noise scale to gate a release of the window. Let $k^{\max} = \arg\max_k M[k]$ denote the global matched-filter argmax. We first estimate the noise scale of $M$ on the current frame by the median absolute deviation,
\begin{align}
\label{eq:sigma_n_mad}
    \sigma_{n,t} \;=\; 1.4826 \cdot \mathrm{median}_k\!\bigl(\bigl|M[k] - \mathrm{median}_{k'} M[k']\bigr|\bigr),
\end{align}
where the factor $1.4826$ is the Gaussian-consistency scaling that makes $\sigma_{n,t}$ a robust estimator of the matched-filter response's standard deviation~\cite{rousseeuw1993alternatives}. The readout abandons the window and switches to the global peak when the global response exceeds the within-window peak by a Shewhart $2\sigma_{n,t}$ absolute margin. The selected integer peak is
\begin{align}
\label{eq:jump_gate}
    \hat k_t \;=\;
    \begin{cases}
        k^{\max},
            & k^{\max} \notin \mathcal{W}_t \text{ and } M[k^{\max}] > \max_{k \in \mathcal{W}_t} M[k] + 2\,\sigma_{n,t}, \\[4pt]
        \hat k_t^{\mathrm{loc}},
            & \text{otherwise.}
    \end{cases}
\end{align}
The additive $2\sigma_{n,t}$ threshold is a two-sigma Shewhart-type absolute margin on the matched-filter response's noise distribution; anchoring it to a runtime MAD estimate makes the gate adapt to the per-frame noise level rather than to a deployment-fixed multiplicative ratio that does not transfer across signal-to-noise regimes.

\paragraph{Adaptive-window MAD-gated centroid sub-bin readout.}
The selected integer peak $\hat k_t$ is refined to sub-bin precision by a gated weighted centroid evaluated on the pooled PSD $S^*_t$ itself, rather than on the matched-filter response $M$. Evaluating the centroid on $S^*_t$ keeps the readout on the sharp pooled sideband instead of the matched-filter-broadened response, and a first-moment (centre-of-mass) estimator over a wide support recovers the \emph{mean} of a sideband that tune motion within a single acquisition has smeared across bins, whereas a vertex estimator locks onto the instantaneous, motion-biased peak top. A reference window of half-width $W_c = \mathrm{round}(q_c / \Delta_q)$ bins around $\hat k_t$ sets the search support, with $q_c = 0.05$ tune units a deployment-fixed window matching the weighted-centroid single-frame baseline; a robust baseline and gate are formed from the window median and median absolute deviation,
\begin{align}
\label{eq:centroid_gate}
    \mathcal{R}_t = \bigl\{\,k : |k - \hat k_t| \le W_c\,\bigr\}, \quad
    m_t = \operatorname{median}_{k\in\mathcal{R}_t} S^*_t[k], \quad
    \tau_t = m_t + 2\,\operatorname{MAD}_{k\in\mathcal{R}_t}\!\bigl(S^*_t[k]\bigr).
\end{align}
The support is then narrowed to the maximal contiguous run $\mathcal{A}_t \subseteq \mathcal{R}_t$ of bins above the baseline $m_t$ that contains $\hat k_t$. This SNR-adaptive support spans the full smeared sideband at high SNR but collapses to a few bins at low SNR, where the noise floor crosses the median within a couple of bins, so it rejects the distant noise spikes that a fixed-width window would admit while still bridging the shallow dips of a bumpy smeared peak. The sub-bin estimate is the gate-weighted centroid over the bins of $\mathcal{A}_t$ that exceed the gate,
\begin{align}
\label{eq:centroid}
    \mathcal{G}_t = \bigl\{\,k \in \mathcal{A}_t : S^*_t[k] > \tau_t\,\bigr\}, \qquad
    \mu_t =
    \begin{cases}
        \dfrac{\sum_{k\in\mathcal{G}_t} k\,\bigl(S^*_t[k] - \tau_t\bigr)}{\sum_{k\in\mathcal{G}_t}\bigl(S^*_t[k] - \tau_t\bigr)}, & |\mathcal{G}_t| \ge 3, \\[10pt]
        \hat k_t, & \text{otherwise.}
    \end{cases}
\end{align}
The integer-peak fallback guards the degenerate low-SNR case in which fewer than three bins pass the gate. The gate factor (two unscaled MADs) and the $(S^*_t - \tau_t)$ weighting are the robust-statistics defaults inherited from the weighted-centroid baseline; the window half-width $q_c$ is the single deployment-tuned parameter of the readout. The continuous-bin estimate $\mu_t$ maps to the reported folded tune by $\hat q_t = \mathrm{clip}(\mu_t,\, 0,\, L - 1)\,\Delta_q$, which lies in $[0, 0.5)$ by construction of the tune-axis grid.

\paragraph{Velocity estimate.}
The motion-compensated coherent exponential moving average (EMA) of this subsection draws its per-frame velocity scalar from the bounded deque of clipped raw velocities derived from the recent tune-output history rather than from the data path. At the start of frame $t$, before the local-window argmax and centroid readout produces $\hat q_t$, the per-frame raw velocity is computed from the two most recent prior outputs as
\begin{align}
\label{eq:v_raw}
    v_{t-1}^{\text{raw}} \;=\; \mathrm{clip}\!\Bigl(\frac{\hat q_{t-1} - \hat q_{t-2}}{\Delta_q},\; -V_{\max},\; +V_{\max}\Bigr),
\end{align}
with $V_{\max} = 2\,W_q$ bins per frame, exactly twice the local-window half-width $W_q$; in tune units this is $2\,w_q = 2 \times 10^{-2}$ per frame ($40$ bins at the deployed $L = 1024$), numerically coinciding with the velocity envelope $v_{\max}$ of the Bayesian tracker of Section~\ref{subsec:nlbf}, so both estimators operate under the same physical velocity prior. The clip bound is nevertheless not a physical-velocity envelope but an algorithmic self-consistency identity: both $\hat q_{t-1}$ and $\hat q_{t-2}$ are local-window argmax outputs each constrained to within $\pm W_q$ of the prior tune anchor, so any unclipped $|v_{t-1}^{\text{raw}}| > 2 W_q$ implies that at least one of the two prior estimates was placed outside its local search window, whether by the global-peak fallback or by a spurious lock; the clip bounds the influence of such a frame pair on the velocity history downstream. The aggregate velocity used during frame $t$ is the arithmetic mean of the most recent $L_v = 3$ clipped samples available at the start of the frame,
\begin{align}
\label{eq:v_hat}
    \hat v_t \;=\; \frac{1}{L_v}\,\sum_{j=1}^{L_v}\,v_{t-j}^{\text{raw}}.
\end{align}
The deployed pipeline uses a three-sample arithmetic mean, an empirical setting that smooths single-frame velocity errors while limiting lag on genuine motion: the raw single-frame velocity propagates transient frame-level errors directly into the alignment, while longer windows lag accelerations. The EMA reads $\hat v_t$ before the current frame's tune is computed, so it uses the velocity going \emph{into} the current frame.

\paragraph{Motion-compensated coherent EMA.}
The matched-filter detector and centroid readout operate on a pooled tune-axis PSD $S^*_t$. Here \emph{coherent} denotes the motion-phase alignment of successive power spectra before averaging, not complex-amplitude (waveform) averaging. A plain exponential moving average of the postprocessed PSDs $\tilde S_t$ is effective only for a nearly stationary tune: under appreciable tune motion, it spreads the historical sideband over the bins visited by the trajectory. We therefore realign the accumulator to the current operating point before each update, translating the previous pooled PSD by the fractional-bin velocity estimate $\hat v_t$ and using a velocity-adaptive decay,
\begin{align}
\label{eq:mc_ema}
    S^*_t \;=\; \beta_t \,\Phi_{\hat v_t}\!\bigl[S^*_{t-1}\bigr] \;+\; (1 - \beta_t)\,\tilde S_t,
    \qquad
    \beta_t \;=\; \min\!\Bigl(\beta_{\max},\; \frac{\sigma_1}{|\hat v_t| + \sigma_1}\Bigr),
    \quad \beta_{\max} = 0.50,
\end{align}
where $\Phi_{\hat v_t}[\,\cdot\,]$ is a fractional-bin translation along the tune axis with linear-interpolation reconstruction and zeroed boundaries, and $\sigma_1$ is the narrowest matched-filter kernel width in bins ($\sigma_1 = 0.5 \times 10^{-3} / \Delta_q \approx 1$ bin at $L = 1024$). The cap $\beta_{\max} = 0.50$ sets the stationary-limit pool depth: under the convention $N = 1 / (1 - \beta)$, which equates the EMA weight on the current frame, $w_0 = 1 - \beta$, to the $1/N$ weight of an $N$-point uniform average, $\beta_{\max} = 0.50$ corresponds to $N = 2$ frames. This depth is a balance between two competing effects at the stationary limit: cross-frame coherent averaging suppresses broadband noise and favours a deeper pool, while residual velocity-estimate error accumulates as read-out lag even at $\hat v_t \approx 0$ (the estimated velocity is never exactly zero) and favours a shallower one. Two frames realise the first and largest increment of the noise-averaging gain while keeping that lag negligible; a third frame ($N = 3$) trades a diminishing noise return against a growing lag, so the cap is set at $N = 2$. As the per-frame tune motion increases, $\beta_t$ drops below the cap once $|\hat v_t| > \sigma_1$, shortening the pool toward the single-frame limit. Although motion compensation removes the bulk shift, residual velocity-estimate error and acceleration accumulate as lag over a deep pool; meanwhile, the wide centroid readout already recovers the within-frame smear, so deeper cross-frame pooling on a fast trajectory adds lag without benefit. No explicit floor is imposed: the velocity clip $|\hat v_t| \le V_{\max}$ bounds the decay below by $\sigma_1 / (V_{\max} + \sigma_1) \approx 0.024$, a ratio of two tune-unit constants ($0.5 \times 10^{-3}$ and $2 \times 10^{-2}$) that is therefore independent of the grid size $L$. The adaptive law engages once the velocity history is populated, from the third frame onward; before then $\hat v_t = 0$ and $\beta_t = \beta_{\max}$. Boundary bins exposed by a non-zero shift are filled with zeros rather than wrapped circularly, preventing spectral content that has moved off the tune axis from re-entering at the opposite edge.

\paragraph{End-to-end frame ordering.}

\begin{figure}
    \centering
    \includegraphics[width=0.95\linewidth]{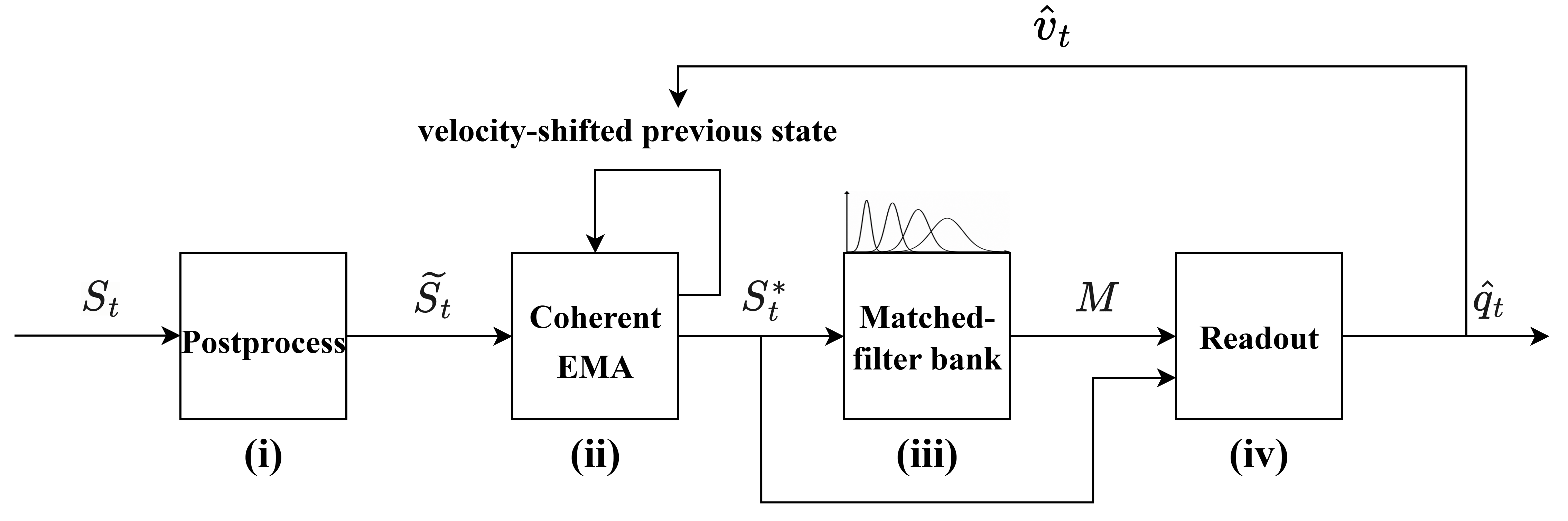}
    \caption{Per-frame pipeline of the classical estimator, comprising the four stages (i)--(iv) detailed in the text. The postprocess block~(i) produces $\tilde S_t$; the motion-compensated coherent EMA~(ii) realigns and pools the postprocessed PSDs into the pooled spectrum $S^*_t$ under a velocity-adaptive decay; the matched-filter bank~(iii) forms the response $M$; and the readout~(iv) locates the integer peak on $M$ and refines it to the sub-bin tune by a centroid evaluated on $S^*_t$ (lower skip path). The unit-delayed velocity feedback $\hat v_t$ drives the realignment shift and the adaptive decay in block~(ii). Frames flagged low-quality by the upstream pipeline bypass the four stages and repeat the previous estimate.}
    \label{fig:emasc_pipeline}
\end{figure}

At each frame $t$, given the prior outputs $\hat q_{t-1}$ and $\hat q_{t-2}$, the deployed pipeline applies four steps in order, summarised in Figure~\ref{fig:emasc_pipeline}. First, the postprocess block above acts on the raw mapped PSD $S_t$ to produce $\tilde S_t$. Second, $v_{t-1}^{\text{raw}}$ is computed from~(\ref{eq:v_raw}), appended to the velocity history, and used to read the updated $\hat v_t$ from~(\ref{eq:v_hat}). Third, the motion-compensated coherent EMA update~(\ref{eq:mc_ema}) uses the updated $\hat v_t$ to produce $S^*_t$. Fourth, the matched-filter bank forms the response $M$ from $S^*_t$, the local-window argmax with Shewhart-gated global fallback selects the integer peak $\hat k_t$ on $M$, and the adaptive-window MAD-gated centroid evaluated on $S^*_t$ refines it to the sub-bin tune $\hat q_t$. A frame whose upstream quality flag marks too few significant peaks bypasses these four steps and repeats the previous estimate. The matched-filter bank and the centroid readout both read the same pooled spectrum $S^*_t$, and the EMA draws its velocity from the readout history only~(\ref{eq:v_hat}), so the design retains a strict feed-forward structure with no estimator-internal cross-coupling between the per-frame readout and the temporal-context buffer beyond the velocity scalar.

\subsection{Deep-learning estimator: likelihood map and discrete tune-velocity Bayes tracker}
\label{subsec:nlbf}
\paragraph{Three-channel feature builder.}\label{para:nlbf-features}
A stateless feature builder converts each $(x, w)$ pair into a three-channel input tensor of shape $(3, L)$, extending the two-channel layout of the dual-branch CNN of~\cite{sun2026real} with an explicit coordinate channel. Channel $c_0$ is a robust per-frame $z$-score of the linear-magnitude folded PSD, computed using the median and median absolute deviation; channel $c_1$ is the soft-binning weight $w[k]$ scaled by its per-frame maximum so that values lie in $[0, 1]$; channel $c_2$ is the folded-tune coordinate $q[k]$ of each bin, which supplies the absolute tune position that the translation-equivariant convolutional trunk cannot otherwise recover. The linear-magnitude representation in $c_0$ is chosen because logarithmic compression reduces the contrast between a low-amplitude transverse sideband and the local noise floor at very low SNR; the $z$-score additionally suppresses pedestal-level offsets between frames and removes the absolute amplitude scale, so the estimator is insensitive to the overall signal level of the acquisition chain. The low-tune mask described in the classical-estimator postprocess block (Section~\ref{subsec:emasc}) zeroes $c_0$ and $c_1$ below the mask edge, while the coordinate channel is left intact. The feature builder carries no learnable parameters and is defined for any spectrum length $L$.

\paragraph{Design rationale.}
The likelihood-map CNN has a deliberately narrow role: given a three-channel input feature tensor of length $L$, it emits per-bin logits $s_t[k]$ that approximate the log-likelihood of the true tune residing at bin $k$, together with a scalar confidence logit $c_t$ that summarises whether the frame is well localised. This contract separates static spectral recognition from temporal inference. The network localises spectral evidence in the presence of noise and clutter, while a discrete state-space tracker described below propagates, smooths, and disambiguates that evidence across frames; the confidence logit is consumed at that stage as a per-frame observation gate. The network therefore operates frame by frame, holds no recurrent state, and is trained on independently sampled frames. The benefit of a likelihood-map output over a scalar regression head is realised at the tracker stage rather than at the single-frame level: the map provides the multi-modal evidence that the downstream Bayes tracker integrates across frames, and the resulting advantage is quantified by the tracker decomposition of Section~\ref{subsec:ablation}.

A further design goal is that the deep-learning estimator should retain the deployment flexibility expected of classical signal processing, which applies to any spectrum length and any absolute signal amplitude without retraining. Amplitude independence is provided by the robust per-frame normalisation of the feature builder above; length generality must come from the network itself, so every architectural component is required to be well defined for an arbitrary spectrum length $L$. Sequence architectures from the language-modelling domain satisfy this requirement while offering the global context that peak discrimination needs: the Transformer~\cite{vaswani2017attention} and the more recent Mamba selective state-space models~\cite{gu2023mamba} both process variable-length inputs natively. Both, however, are substantially heavier than this task requires: quadratic-cost attention in the former and input-dependent selective recurrence in the latter are built for general sequence modelling, whereas this front end performs static per-frame peak localisation and delegates temporal aggregation to the tracker. The trunk therefore adopts the core computational primitive shared by the linear-time-invariant state-space predecessors of Mamba~\cite{gu2022parameterization} and their long-convolution successors~\cite{poli2023hyena}: a global convolution with a compact, implicitly parameterised kernel, evaluated exactly with the fast Fourier transform.

The principal architectural choices follow three constraints inherited from this contract: per-frame inference latency on commodity GPUs, deployment under variable batch size and variable spectrum length, and a multi-modal output that the downstream Bayes tracker can ingest directly. Three consequences follow.

\emph{Trunk.}
The trunk pairs short local convolutions with the FFT-evaluated global convolution motivated above. The local branch resolves fine spectral structure at the bin scale, while the global branch provides a gapless receptive field over the entire tune axis at $O(L \log L)$ cost, so distant context, such as the absence of competing peaks elsewhere on the axis, informs every bin without recurrence or attention. The stack remains purely feed-forward at inference. Recurrent or attention-based trunks were not selected here because temporal aggregation is delegated to the tracker, which reduces their expected benefit for this per-frame front end and would add per-frame cost.

\emph{Normalisation and stochasticity.}
The trunk uses GroupNorm rather than BatchNorm. The deployed estimator processes one frame at a time, so any normalisation that depends on batch-level statistics would require inference-time running estimates and re-introduce a coupling between training and deployment statistics. GroupNorm is computed entirely from within-frame activations and is batch-size independent. The architecture contains no dropout, which preserves deterministic inference; robustness to narrow-band contaminants is handled explicitly by the training augmentation described in Section~\ref{subsec:setup}.

\emph{Map supervision and head separation.}
The map head is supervised with a localised negative log-likelihood evaluated inside a fixed, physics-derived support window centred at the continuous, sub-bin true-tune position. The per-frame peak width is self-measured from the map itself rather than imposed by a target template. This keeps the sub-bin offset of the tune in the gradient at every spectrum length and avoids dictating a peak shape that in reality varies with the upstream chain configuration; the corresponding loss is detailed in the training-objective formulation below. The map and confidence heads are kept parallel rather than serial so that the confidence logit does not feed back into the map output. Both heads share the trunk feature so that the same representation underlies the likelihood map and the frame-quality prediction that gates the tracker.

\begin{figure}
    \centering
    \includegraphics[width=0.9\linewidth]{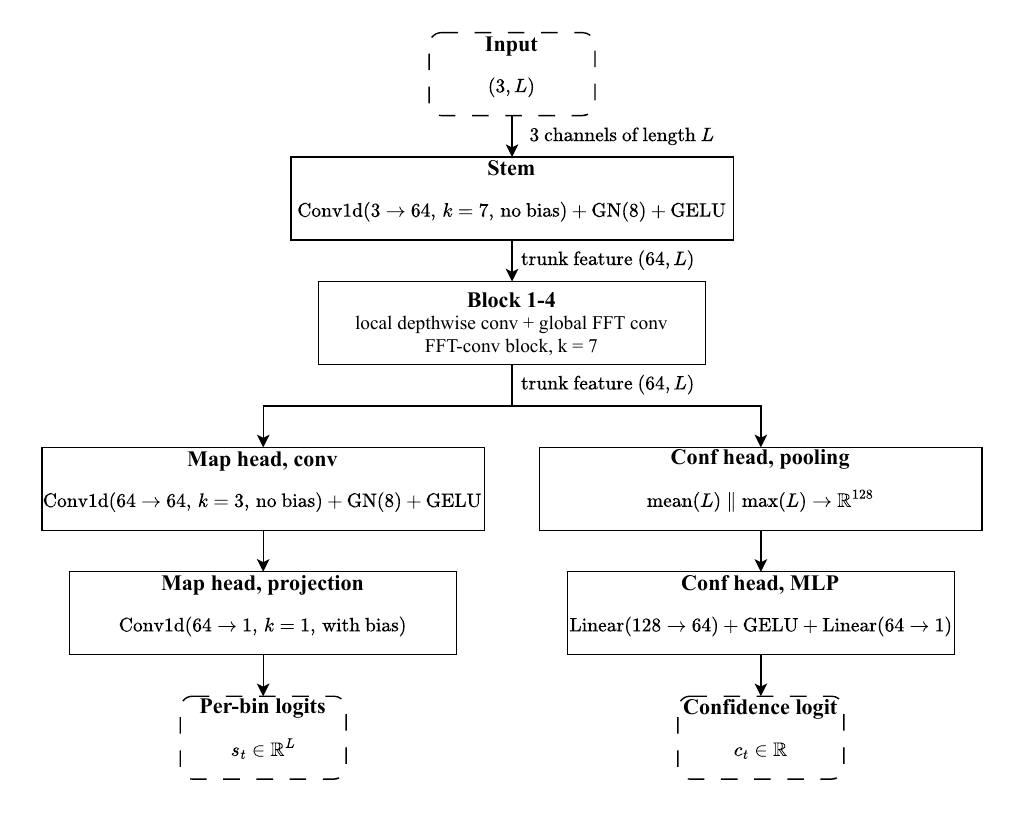}
    \caption{Overall topology of the likelihood-map CNN. The three-channel input feature tensor of length $L$ is processed by a convolutional stem, a trunk of four FFT-augmented convolutional blocks, and two parallel output heads. The map head emits per-bin logits over the $L$-bin tune grid; the confidence head emits a scalar logit whose sigmoid gates the tracker observation. Channel widths, kernel sizes, and the global-convolution parameterisation are described in the prose.}
    \label{fig:cnn_overall}
\end{figure}

\begin{figure}
    \centering
    \includegraphics[width=0.6\linewidth]{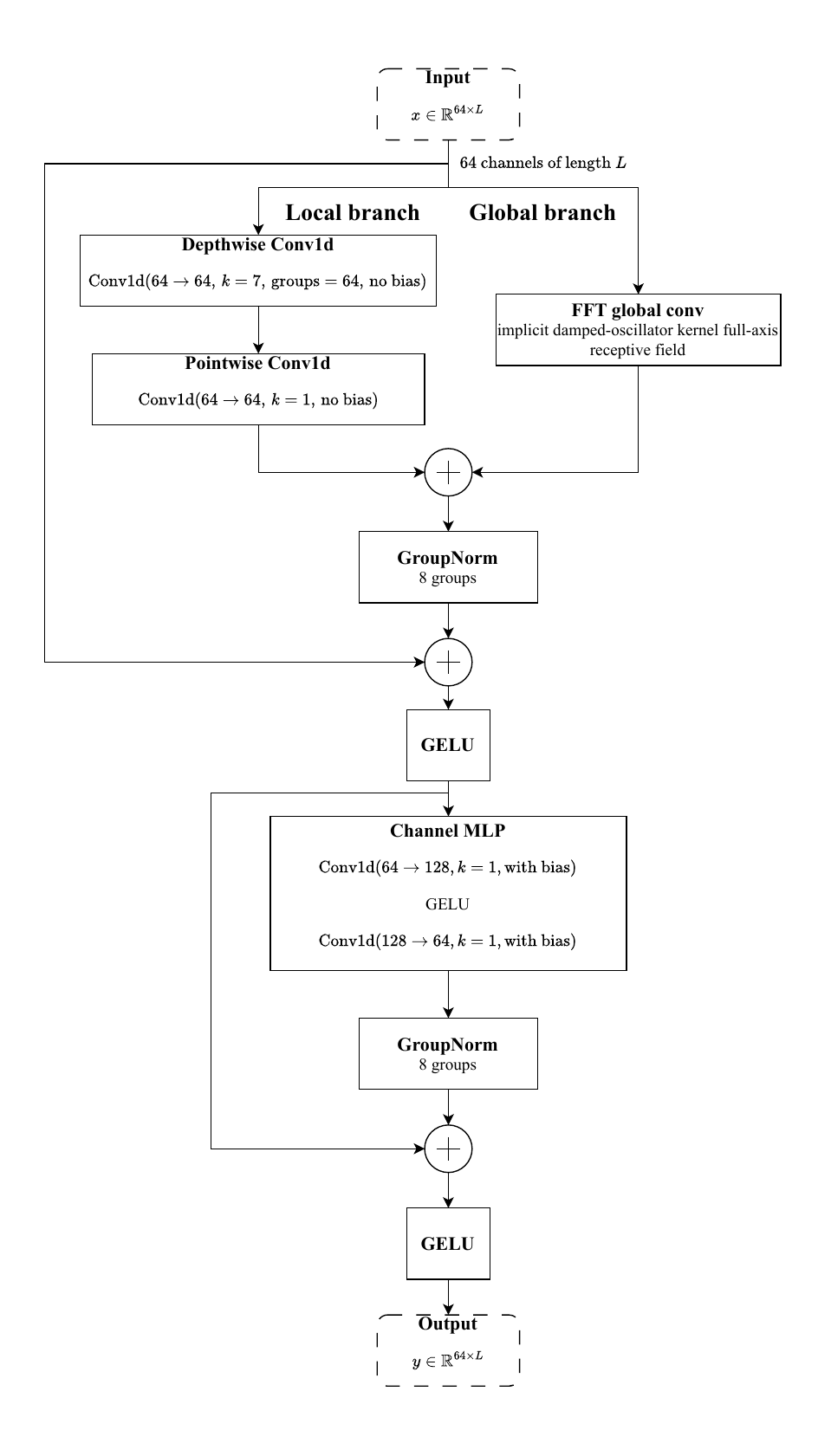}
    \caption{Trunk block geometry. Each block sums a local branch (depthwise kernel-size-$7$ convolution followed by a pointwise $1 \times 1$ convolution) and a global branch (FFT-evaluated global convolution with an implicitly parameterised kernel spanning the full tune axis), followed by GroupNorm, a residual connection, and a GELU; a channel MLP with its own GroupNorm and residual connection completes the block.}
    \label{fig:cnn_block}
\end{figure}

\paragraph{Architecture overview.}
The network contains approximately $109\,000$ trainable parameters and is organised into three stages: a stem, a trunk of four FFT-augmented convolutional blocks, and two parallel output heads. Figure~\ref{fig:cnn_overall} shows the overall topology, and Figure~\ref{fig:cnn_block} shows the geometry of one block. The stem maps the three input channels to $64$ channels with a single kernel-size-$7$ 1D convolution, followed by GroupNorm and a Gaussian Error Linear Unit (GELU) activation~\cite{gelu}; the kernel size of $7$ used in the stem and in the local branch below follows the kernel-size ablation of the regression-CNN backbone of~\cite{sun2026real}. Each trunk block mixes the $64$-channel feature along the tune axis through two parallel branches: a local branch, a depthwise kernel-size-$7$ convolution followed by a pointwise $1 \times 1$ convolution, and the global branch described next. The two branch outputs are summed, normalised with GroupNorm, added to the block input through a residual connection, and passed through a GELU activation; a channel multilayer perceptron (MLP), formed by two $1 \times 1$ convolutions expanding $64 \to 128 \to 64$ with a GELU between them, with its own GroupNorm and residual connection, completes the block.

The global branch realises, for each channel $c$, a convolution spanning the full tune axis whose kernel is not stored as $2L - 1$ free taps but generated from five scalars as a damped-oscillator profile over the normalised relative position $p \in [-1, 1]$,
\begin{align}
\label{eq:fftkernel}
    K_c(p) \;=\; w_{1,c}\, e^{-s_{1,c}\lvert p \rvert} \cos\!\bigl(\pi f_c\, p\bigr)
    \;+\; w_{2,c}\, e^{-s_{2,c}\lvert p \rvert},
\end{align}
with amplitudes $w_{1,c}, w_{2,c}$, positive decay rates $s_{1,c}, s_{2,c}$, and oscillation frequency $f_c$, sampled on the $2L - 1$ relative offsets of the frame. The convolution is evaluated exactly (zero-padded to length $3L$, hence linear rather than circular) in the Fourier domain at $O(L \log L)$ cost, and the amplitudes are initialised near zero so that the global branch starts near-silent and grows during training only where global context helps. Because $p$ is normalised and every frame spans the same folded-tune interval $[0, 0.5)$, the kernel's physical shape in tune units is independent of $L$ by construction: the same trained network applies unchanged to any spectrum length, in contrast to a dilated-convolution trunk whose receptive field is fixed in bins and therefore shrinks in tune units as $L$ grows. No component of the network has a fixed-length dependence, so a single set of weights serves all evaluated grids, with $L = 1024$ the deployed SAPT operating point.

The \emph{map head} reduces the trunk feature to the per-bin logits $s_t \in \mathbb{R}^{L}$ via a kernel-size-$3$ convolution with GroupNorm and a GELU activation, followed by a $1 \times 1$ convolution that maps the $64$-channel feature to a single output channel. No softmax is applied at this stage; the training loss reads the logits through a temperature-$0.5$ softmax for the windowed map term, while the tracker masks invalid bins and applies a unit-temperature $\log\mathrm{softmax}$ to form the per-frame observation. The \emph{confidence head} pools the trunk feature by concatenating its channel-wise mean and channel-wise maximum over the tune axis, and passes the pooled $128$-dimensional vector through a small multilayer perceptron with one hidden layer of $64$ units, a GELU activation, and a single scalar output. The output $c_t$ is the pre-sigmoid logit for the event that the map argmax lies within twice the self-measured peak width of the continuous true-tune position, as defined with the training objective below. Unlike an auxiliary head that regularises training and is discarded afterwards, the confidence head is retained at inference: the deployed estimator converts it to the per-frame observation weight $w_{\mathrm{obs},t} = \mathrm{sig}(c_t)$ that gates the tracker update described below.

\paragraph{Training objective.}
The objective has three active training signals: a windowed localisation loss on the map head that teaches the network where the sideband evidence should lie, a peak-dominance term that ranks the true peak above the rest of the axis, and a binary cross-entropy loss on the confidence head that calibrates the frame-quality probability consumed by the tracker gate. Concretely,
\begin{align}
\label{eq:loss}
    \mathcal{L} = \mathcal{L}_{\mathrm{map}}
                + w_{\mathrm{conf}}\,\mathcal{L}_{\mathrm{conf}}
                + w_{\mathrm{dom}}\,\mathcal{L}_{\mathrm{dom}},
\end{align}
with $w_{\mathrm{dom}} = 0.1$ and $w_{\mathrm{conf}} = \ln 2 / \ln L$, the ratio of the natural entropy scales of the two supervision signals (a binary cross-entropy saturates at $\ln 2$\,nats for an uninformative prediction, a distribution over $L$ bins at $\ln L$\,nats); at the deployed $L = 1024$ the schedule gives $w_{\mathrm{conf}} = 0.1$. Because the confidence target is computed from the map argmax, which is effectively random at initialisation, the small relative weight also limits the influence of the confidence-loss gradient on the shared trunk until the map output has stabilised. The three terms are detailed below.

\emph{Windowed map loss.} Let $k^*_{\text{frac}} \in [0, L)$ be the continuous bin position of the ground-truth tune. The map is read as a probability vector at the deployed readout temperature, $r_t = \mathrm{softmax}(s_t / \tau)$ with $\tau = 0.5$\,bin, so that training, checkpoint selection, and the deployed single-frame readout all evaluate the same distribution. A fixed support window $a[k] \in [0, 1]$ is centred at the truth: $a = 1$ within a core radius of $\sigma_{\max} / \Delta_q$ bins, decaying smoothly (cosine taper) to $0$ at $2\sigma_{\max} / \Delta_q$ bins, where $\sigma_{\max} = 0.025$ tune units is an upper bound on the bunched-beam sideband width and $\Delta_q = 0.5/L$ the bin width, so the window is fixed in tune units at every $L$. The window encodes a maximal-support assumption (a real sideband fits inside it) rather than a target width: it excludes distant distractors from the localisation gradient, and because it is anchored at the label rather than at any map quantity, it acts as a constant mask with respect to the logits. Within the window, the loss measures the map against itself: with in-window mass $m = \sum_k r_t[k]\, a[k]$ and localised distribution $\rho = r_t\, a / m$, the localised mean $\hat k = \sum_k \rho[k]\, k$ and spread $\hat\sigma = \bigl(\sum_k \rho[k]\,(k - \hat k)^2\bigr)^{1/2}$ (floored at $0.5$\,bin) form a Laplace negative log-likelihood, complemented by a barrier on the window mass,
\begin{align}
\label{eq:loss_map}
    \mathcal{L}_{\mathrm{map}}
        = \frac{\lvert \hat k - k^*_{\text{frac}} \rvert}{\hat\sigma} + \log \hat\sigma
        + \Bigl[\max\Bigl(0,\, \log\frac{m_{\min}}{m}\Bigr)\Bigr]^{2},
    \qquad
    m_{\min} = \frac{2 \cdot 2\sigma_{\max}}{0.5} = 0.20,
\end{align}
where $m_{\min}$ is the fraction of the folded-tune axis spanned by the window's full outer extent; because the cosine taper admits less than this fraction of a flat, no-peak map, the barrier requires the network to hold strictly more probability near the truth than an uninformative spectrum would, while leaving up to $80\%$ of the probability mass free for legitimate secondary peaks elsewhere on the axis. Because the width $\hat\sigma$ is self-measured per frame, no peak template and no signal-to-noise oracle enter the supervision, and the sub-bin offset of the tune remains in the gradient through $\hat k$. Every training frame carries a transverse sideband, so this localisation objective is applied to all frames.

\emph{Peak dominance.} The windowed loss constrains the map only inside the support window, so a distractor may still outrank the true peak elsewhere on the axis; this matters for the single-frame readout even though the tracker disambiguates such frames through its predicted position. The dominance term is the contrastive log-odds of the true bin against all out-of-window bins,
\begin{align}
\label{eq:loss_dom}
    \mathcal{L}_{\mathrm{dom}}
        = -\, m \, \log
        \frac{e^{s_t[k^*]}}{e^{s_t[k^*]} + \sum_{k \notin \mathcal{W}} e^{s_t[k]}},
\end{align}
where $k^*$ is the integer true bin, $\mathcal{W}$ is the support window, and the in-window mass $m$ continuously down-weights the term according to the probability the map places near the truth, so the dominance pressure fades on frames too noisy to localise. Because in-window bins other than $k^*$ are excluded from the denominator, the term raises the peak above distant distractors without sharpening it against its own shoulders.

\emph{Confidence calibration.} The confidence head is trained as a per-frame frame-quality classifier whose calibrated probability the deployed tracker consumes as its observation gate. Its target is derived from the current map output (the frame's argmax outcome) and is detached during back-propagation, so the confidence loss does not supervise the map head through this self-generated label; the gradient updates the confidence head and the shared trunk. The label is one when the network's argmax lies within twice the self-measured spread of the truth, and zero otherwise. Using binary cross-entropy (BCE) on the confidence logit, the loss is
\begin{align}
\label{eq:loss_conf}
    \mathcal{L}_{\mathrm{conf}}
        = \mathrm{BCE}\!\bigl(c_t,\, \mathbb{1}[\,\lvert \arg\max_k s_t[k] - k^*_{\text{frac}} \rvert < 2\hat\sigma\,]\bigr),
\end{align}
where $\mathrm{BCE}$ is applied to the sigmoid of the confidence logit $c_t$ and the radius $2\hat\sigma$ reuses the localised spread of the windowed map loss, so the correctness criterion adapts to the peak width the map itself reports. The BCE supervision trains $\mathrm{sig}(c_t) \in [0, 1]$ to estimate the probability that the current frame's argmax lies within $2\hat\sigma$ of the truth, and the tracker applies this probability as the per-frame observation weight.

\begin{figure}
    \centering
    \includegraphics[width=0.9\linewidth]{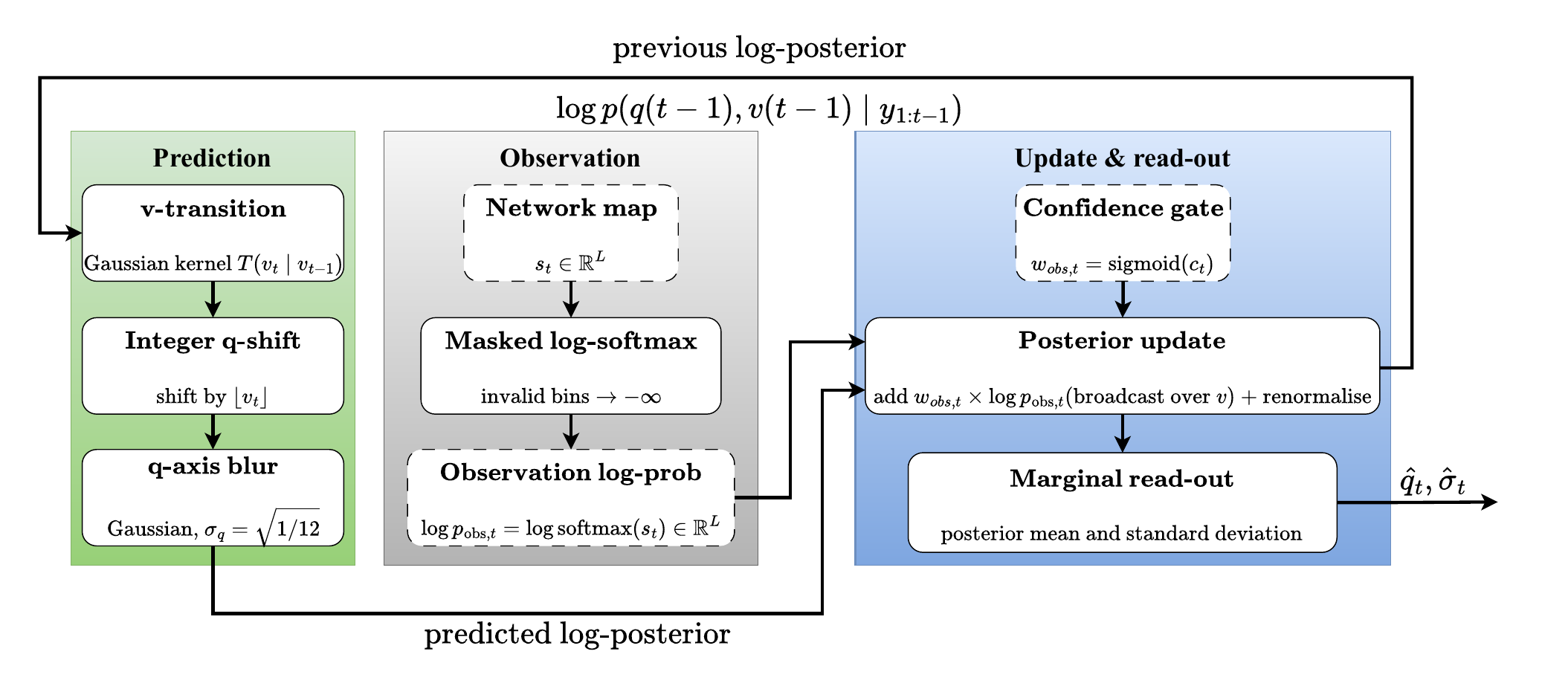}
    \caption{Discrete $(q, v)$ Bayes tracker recursion. The three panels correspond to the per-frame \emph{prediction}, \emph{observation}, and \emph{update \& read-out} stages detailed in the text. Dashed-border rectangles denote a panel's per-frame external inputs or outputs; solid-border rectangles denote operations performed inside a panel. The confidence gate $w_{\mathrm{obs},t} = \mathrm{sig}(c_t)$ enters the update panel as a per-frame external input that scales the observation log-likelihood. The top arrow forwards the full $(q, v)$ log-posterior tensor losslessly into the next frame's prediction stage, in parallel with the marginal read-out that projects it to the reported scalar pair $(\hat q_t, \hat \sigma_t)$.}
    \label{fig:tracker_pipeline}
\end{figure}

\paragraph{Discrete $(q, v)$ Bayes tracker.}
At inference time the trained network produces, for each incoming frame, the per-bin logits $s_t \in \mathbb{R}^{L}$ and the confidence logit $c_t$; the tracker consumes the former as its per-frame observation and the sigmoid of the latter as the weight of that observation. The tracker propagates a log-domain posterior over the internal $(k, v)$ grid, $\boldsymbol{\Lambda}_t \in \mathbb{R}^{L \times (2V_{\max}+1)}$, whose tune axis coincides with the $L$-bin spectrum grid. All motion parameters are specified in physical tune units and converted to bins at the operating $L$, so the tracker, like the network, is defined once for all spectrum lengths. The velocity envelope is $v_{\max} = 2 \times 10^{-2}$ tune units per frame, giving $V_{\max} = \mathrm{round}(v_{\max} / \Delta_q) = 41$ bins per frame at the deployed $L = 1024$; at the millisecond-scale per-frame acquisition window typical of SAPT operation, this envelope covers the slow-extraction and acceleration tune motion encountered at compact medical proton synchrotrons, including the fastest ramp trajectories evaluated in Section~\ref{sec:results}, and leaves headroom for use under other operating conditions or at other facilities without changing the tracker architecture. Each per-frame step comprises three stages, executed in the order observation, update plus read-out, and prediction, the last preparing the prior for the next frame, as summarised in Figure~\ref{fig:tracker_pipeline}.

\emph{Prediction.}
The joint $(q, v)$ transition density is applied in two factored steps to avoid materialising a transition operator on the full joint grid: the state memory stays proportional to $L \cdot (2V_{\max} + 1)$, the velocity transition multiplies each tune bin's velocity row by a $(2V_{\max} + 1) \times (2V_{\max} + 1)$ kernel, and the subsequent tune-axis shift and blur are linear in the state size. First, a velocity transition kernel
\begin{align}
\label{eq:v-kernel}
    T(v_t \mid v_{t-1}) \propto
    \exp\!\bigl(-(v_t - \rho\, v_{t-1})^2 / (2\sigma_v^2)\bigr)
\end{align}
propagates mass across the velocity axis at fixed tune bin. The tracker uses $\rho = 1.0$, so the velocity transition is a random walk that encodes the design prior of persistent per-frame tune velocity. The process-noise scale is a physical, facility-level calibration, $\sigma_v = 5 \times 10^{-4}$ tune units per frame (approximately one bin per frame at $L = 1024$), set so that $4\sigma_v$ covers the largest per-frame velocity change measured on the evaluated trajectories; it is a property of the beam, not of the spectrum resolution, and is therefore not scaled with $L$. Since a velocity spread finer than the grid cannot diffuse between velocity states, the bin-domain value of $\sigma_v$ is floored at one velocity-grid step; the floor binds only when the bin width $0.5/L$ exceeds $\sigma_v$, i.e. for $L < 1000$, which among the power-of-two grids evaluated in this work is only $L = 512$. Second, for each velocity bin $v$, the tune-bin axis is shifted by the corresponding integer number of bins, vectorised across the velocity dimension. A one-dimensional Gaussian convolution with standard deviation $\sigma_q = 1/\sqrt{12} \approx 0.289$\,bins is then applied along the tune-bin axis to model the residual sub-bin uncertainty introduced by the integer shift; the variance $\sigma_q^2 = 1/12$ matches that of a uniform quantisation remainder on $[0, 1)$. The convolution width is the floor-residual variance only and does not include an additive $\sigma_v^2$ term: the velocity-process noise $\sigma_v$ is already propagated into the joint $(q, v)$ distribution by the velocity-transition kernel of~(\ref{eq:v-kernel}) and the subsequent integer q-shift, so an extra $\sigma_v^2$ contribution inside the q-blur would double-count the velocity noise on this two-stage discrete topology. For deployments where the tracker update dominates the latency budget, the velocity axis can be subsampled by an integer stride at a fixed velocity range, in which case the quantisation blur widens in proportion to the stride, one velocity-lattice quantum; all configurations evaluated in this work use the native stride-one grid.

\emph{Observation.}
Per frame, the observation log-likelihood is the masked log-softmax of the network map logits $s_t \in \mathbb{R}^{L}$,
\begin{align}
\label{eq:obs-likelihood}
    \log p_{\mathrm{obs},t}[k] = \log\mathrm{softmax}\bigl(s_t\bigr)[k],
\end{align}
where bins outside the valid-mask subset $\mathcal{V}$ are masked to $-\infty$ in $s_t$ before the softmax, so they receive zero posterior mass without requiring a separate clutter-floor term. The normalisation is taken at unit temperature, so the tracker sees a slightly blunter likelihood than the sharpened ($\tau = 0.5$) single-frame readout; the blunter form is deliberately conservative, avoiding over-commitment to any single frame before temporal evidence has accumulated. The learned likelihood map is the estimator's sole per-frame observation, and the confidence gate below sets how strongly it is applied.

\emph{Update and read-out.}
The log-posterior is updated by adding the observation log-likelihood, scaled by the confidence gate and broadcast across the velocity dimension, and renormalising,
\begin{align}
\label{eq:gate-update}
    \boldsymbol{\Lambda}_t[k, v] \;\leftarrow\; \boldsymbol{\Lambda}_t[k, v]
    \;+\; w_{\mathrm{obs},t}\, \log p_{\mathrm{obs},t}[k],
    \qquad
    w_{\mathrm{obs},t} = \mathrm{sig}(c_t),
\end{align}
which tempers the per-frame likelihood with the network's own calibrated reliability: the updated posterior is proportional to the prior times $p_{\mathrm{obs},t}^{\,w_{\mathrm{obs},t}}$, so a confident frame applies close to the full Bayes update, while a frame the network reports as unreliable, such as a beam-loss gap or a noise-dominated spectrum, contributes an almost flat likelihood and the filter coasts on its motion model instead of locking onto a noise peak. The gate is a single multiplication with no data-dependent branching, so the per-frame recursion remains a fixed dense computation. The reported point estimate and standard deviation are computed from the marginal bin distribution
\begin{align}
\label{eq:readout-marginal}
    p_t[k] = \sum_v p(k_t = k, v_t = v \mid \mathbf{y}_{1:t})
\end{align}
as
\begin{align}
\label{eq:readout}
    \hat k_t &= \mathbb{E}_{p_t}[k], &
    \hat \sigma_{k,t} &= \sqrt{\mathrm{Var}_{p_t}[k]}, \\
    \hat q_t &= \hat k_t\,\frac{0.5}{L}, &
    \hat \sigma_t &= \hat \sigma_{k,t}\,\frac{0.5}{L}.
\end{align}
Consequently, $\hat \sigma_t$ widens when the observation evidence is ambiguous, and also while the gate suppresses a run of unreliable frames, because it is computed from the marginal discrete-state posterior rather than predicted by a learned aleatoric-uncertainty head as in~\cite{sun2026real}.

\section{Experimental results}
\label{sec:results}
We evaluate the two proposed estimators of Section~\ref{sec:methods} alongside five published or commonly used baselines on the synthetic studies and operational real-beam data. The seven estimators are the classical estimator of Section~\ref{subsec:emasc}; the deep-learning estimator of Section~\ref{subsec:nlbf}; the dual-branch-CNN-with-Kalman-filter pipeline of~\cite{sun2026real}, referred to as CNN+KF; the temporal peak-detection algorithm introduced in~\cite{sun2025high}, hereafter T-PD; and three stateless single-frame baselines drawn from classical Schottky-tune analysis~\cite{betz2017bunched,lasocha2022estimation}, namely peak detection (PD), curve fitting (CF), and weighted centroid (WC). All estimators are reported under their deployed configuration, with a single value per cell across the benchmarks of this section. Sections~\ref{subsec:setup} to~\ref{subsec:ablation} report the synthetic studies and the ablations, Section~\ref{subsec:real_beam} presents the preliminary validation on operational beam data, and Section~\ref{subsec:decision_matrix} presents the operating-regime decision matrix that maps operating regime and deployment constraints to a recommended estimator.

\subsection{Evaluation setup}
\label{subsec:setup}
\paragraph{Training data.}
The dataset spans $-20$ to $0$\,dB SNR and covers operational ranges of revolution frequency, detector bandwidth, sampling rate, and beam-distribution parameters representative of the SAPT facility, with the betatron tune sampled on the folded range used by the companion preprocessing. The full unique-frame set comprises $250\,000$ frames and is partitioned in a $3:1:1$ ratio into $150\,000$ training, $50\,000$ validation, and $50\,000$ test frames; no new Schottky-signal generator is introduced for this work~\cite{sun2026real}.

\paragraph{Augmentation.}
With probability $0.21$, a training frame receives one, two, or three spurious Gaussian peaks with equal conditional probability; otherwise it is left unchanged. When a per-frame target tune is available, each injected peak is placed with probability $0.5$ in the hard-negative band at a separation of $H$ to $3H$ from the true peak (with $H$ fixed in tune units at approximately $3 \times 10^{-3}$ and rounded to the operating grid, i.e.\ $6$ bins at the deployed $L = 1024$: an inner edge of about two sideband linewidths at every spectrum length, close enough to the true sideband to be confusable but far enough to remain a resolvable distinct peak), and otherwise at a uniformly random tune position above the low-tune cut; amplitudes are randomised within calibrated ranges and Gaussian widths span $\sigma \in [0.5, 1.27]$ bins ($\sigma_{\max}$ derived from a $3$-bin maximum FWHM via the Gaussian identity $\sigma = \mathrm{FWHM}/(2\sqrt{2 \ln 2})$; the width range is kept in bins deliberately, as a resolution-limited narrow-band contaminant occupies a near-constant number of bins at any spectrum length). This augmentation exposes the network during training to narrow-band contaminants whose amplitudes range from comparable to several times larger than those of the transverse betatron sidebands.

\paragraph{Training hyper-parameters for the deep-learning estimator.}
The likelihood-map CNN described in Section~\ref{subsec:nlbf} is trained for $30$ epochs with the AdamW optimiser~\cite{adamw}, using a learning rate of $3 \times 10^{-4}$, weight decay of $10^{-4}$, and a one-epoch linear warmup followed by cosine decay to zero; the deployed checkpoint is selected on validation metrics.

\paragraph{Cross-fold protocol for the deep-learning estimator.}
Five deep-learning checkpoints are trained under the architecturally identical configuration of Section~\ref{subsec:nlbf} and identical training hyper-parameters in a five-fold cross-validation: the $250\,000$-frame set is partitioned into five equal groups and each checkpoint holds out a different group as its test set, training on the remaining four (a $150\,000$/$50\,000$/$50\,000$ train/validation/test split per fold). One of these checkpoints is deployed for evaluation; within each fold the best epoch is selected on validation metrics, never on test-set performance. The deep-learning-estimator numbers reported in Sections~\ref{subsec:static} to~\ref{subsec:ablation} are obtained from that single deployed checkpoint, which Section~\ref{subsec:ablation} confirms is representative across all five folds rather than a favourable-split selection.

\paragraph{Main dynamic-tune benchmark.}
The main synthetic evaluation domain is a $3 \times 5$ grid of synthetic dynamic-tune sequences generated from controlled-SNR Schottky simulations under broadband noise only. The grid spans the two operationally dominant axes identified in Section~\ref{sec:intro}: trajectory speed and per-frame SNR. The three trajectory shapes, referred to as \emph{slow}, \emph{medium}, and \emph{fast}, are ordered by increasing per-frame tune-step magnitude. The mean and maximum per-frame tune steps are approximately $(0.6, 1.7) \times 10^{-3}$ tune units for slow, $(1.6, 4.2) \times 10^{-3}$ for medium, and $(2.5, 9.4) \times 10^{-3}$ for fast, corresponding on the default $L = 1024$-bin grid (the deployed setting, used by every benchmark in this section except the resolution sweeps of Section~\ref{subsec:ablation}) to approximately $(1.2, 3.4)$, $(3.2, 8.6)$, and $(5.1, 19.3)$ bins per frame, respectively. The five SNR levels are $0$, $-5$, $-10$, $-15$, and $-20$\,dB. Within each of the $15$ cells the full estimator pipeline is run on the full sequence and the metric is computed after the common $30$-frame tracker warm-up exclusion. The reported metric is the mean absolute tune error $\mathrm{MAE}_q = \mathbb{E}[\lvert\hat q_t - q_t\rvert]$, with all numerical values quoted in units of $10^{-3}$.

\paragraph{Baselines.}
The CNN+KF baseline~\cite{sun2026real} pairs a lightweight dual-branch CNN with calibrated aleatoric uncertainty with a one-dimensional uncertainty-aware Kalman filter. Because the T-PD baseline~\cite{sun2025high} is non-realtime by design, we apply a group-delay latency compensation to its per-frame output: the integer frame shift $N^{*}$ is selected per trajectory shape as the value that minimises T-PD's mean absolute error over that shape's five SNR levels (per shape, not per SNR); on the evaluated benchmark generation all three shapes select $N^{*} = 3$\,frames, consistent with the EMA group delay being a property of the filter cascade rather than of the trajectory. The selected $N^{*}$ is applied in every cell of every benchmark, so T-PD is evaluated at its best group-delay alignment rather than with an uncompensated readout. The three stateless single-frame baselines~\cite{betz2017bunched,lasocha2022estimation} are \emph{peak detection} (argmax of a Gaussian-smoothed spectrum within the valid tune-grid range), \emph{curve fitting} (Gaussian peak with linear background, fit to a window around the coarse peak with bounded amplitude and width parameters), and \emph{weighted centroid} (gated centre-of-mass within a window around the coarse peak, with a robust amplitude threshold suppressing low-significance bins); they have no temporal memory, and their implementations follow the static comparison of~\cite{sun2026real}.

The five baselines and the two proposed estimators receive identical input frame streams through the shared upstream preprocessing pipeline of Section~\ref{subsec:preprocessing} and are evaluated on identical sequences, so the reported differences reflect estimator design rather than input construction.

\paragraph{SNR convention.}
The SNR labelling throughout this section follows the per-band power convention of Section~\ref{subsec:schottky}; finite-particle fluctuations in the realised per-frame sideband power are absorbed into the reported metric variance and not separately reported.

\subsection{Static, single-frame benchmark}
\label{subsec:static}
The static comparison isolates the spectral-recognition layer of the deep-learning estimator by evaluating it frame by frame, without temporal context, against the regression head of the CNN+KF baseline of~\cite{sun2026real} and the three stateless single-frame baselines PD, CF, and WC. The deep-learning estimator's likelihood-map output is reduced to a single per-frame tune estimate by a peak-local soft-argmax: a hard argmax over the low-cut-masked $L = 1024$-bin map selects the dominant peak, and the expectation $\hat q = \sum_{k} \mathrm{softmax}(s_t/\tau)[k]\,q_{\mathrm{grid}}[k]$ ($\tau = 0.5$, the deployed readout temperature) is evaluated within the training window's outer radius of that peak. The local readout is the fair single-frame reduction for this network: the windowed training objective of Section~\ref{subsec:nlbf} constrains the map only near the true tune and deliberately leaves the far-field mass free, so a whole-map centroid would be penalised by construction rather than by recognition quality. This setup isolates the architectural difference between a likelihood-map output and a scalar regression head at the frame level.

\begin{table}[htbp]
\centering
\caption{Static single-frame benchmark: cross-architecture frame-level comparison on the $50\,000$-frame i.i.d.\ test split (SNR $\in [-20, 0]$\,dB, no temporal state). All values are in $10^{-3}$ tune units except the catastrophic-miss column, which is the percentage of frames whose absolute error exceeds $5 \times 10^{-3}$. ``deep-learning CNN-only'' is the deployed likelihood-map CNN of the deep-learning estimator with the peak-local soft-argmax readout ($\tau = 0.5$) described in the prose; ``CNN+KF CNN-only'' is the regression head of the CNN+KF baseline of~\cite{sun2026real} with the Kalman filter disabled. Bold marks the smallest value per column among the five estimators. \label{tab:static}}
\smallskip
\footnotesize
\begin{tabular}{l c c c c c}
\toprule
estimator & MAE & p50 & p95 & p99 & $T_{5 \cdot 10^{-3}}$ (\%) \\
\midrule
deep-learning CNN-only & 2.529 & 0.114 & 2.719 & 65.506 & 2.780 \\
CNN+KF CNN-only      & \textbf{1.193} & \textbf{0.075} & \textbf{1.912} & \textbf{10.903} & \textbf{1.962} \\
PD                   & 13.581 & 0.277 & 102.746 & 295.706 & 11.960 \\
CF                   & 13.190 & 0.170 & 103.187 & 297.103 & 9.334 \\
WC                   & 13.895 & 0.603 & 101.975 & 293.631 & 15.596 \\
\bottomrule
\end{tabular}
\end{table}

\begin{figure}[htbp]
\centering
\includegraphics[width=0.85\textwidth]{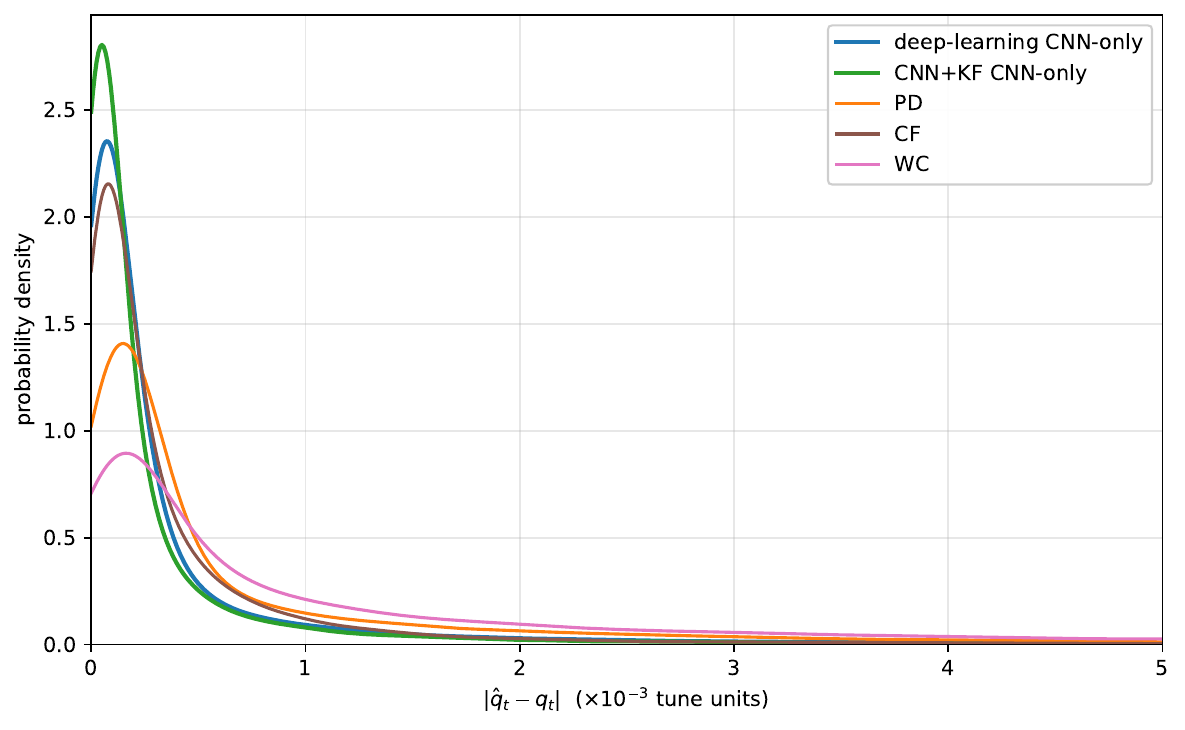}
\caption{Probability density of the per-frame absolute tune error $|\hat q_t - q_t|$ on the i.i.d.\ test split of Table~\ref{tab:static}, estimated by Gaussian kernel density over the body of the distribution ($|\hat q_t - q_t| \le 8 \times 10^{-3}$ tune units); the heavier tail is summarised by the $p99$ column of Table~\ref{tab:static} rather than reproduced here. \label{fig:static_envelope}}
\end{figure}

Table~\ref{tab:static} reports this frame-level comparison. At the single-frame level the scalar regression head is now the stronger reader: the CNN+KF head holds every column. The gap is a tail phenomenon rather than a central one, with the two CNN estimators close at the median but diverging in the upper percentiles, because on ambiguous deep-SNR frames the likelihood map, trained to represent per-frame uncertainty rather than to commit to a point, places mass on more than one candidate and its argmax lands on the wrong peak more often than the regression head's pooled scalar. Both CNN frame-level estimators nevertheless remain well below the three stateless baselines in MAE. Figure~\ref{fig:static_envelope} visualises the per-frame absolute-error distributions as smoothed envelopes, where the two CNN estimators show visibly sharper concentration near zero error than the three stateless baselines. This frame-level outcome sharpens the design hedge of Section~\ref{subsec:nlbf}: the likelihood-map output is \emph{less} accurate than scalar regression as a stand-alone single-frame estimator, and its value lies in what it feeds downstream; the per-frame ambiguity that costs it this benchmark is exactly the information the discrete $(q, v)$ Bayes tracker consumes and resolves across frames. The substantial dynamic-regime separation between the deep-learning estimator and the CNN+KF baseline reported in the next subsection, and the component decomposition of Section~\ref{subsec:ablation}, therefore attribute the deployed estimator's accuracy to the map-plus-tracker pair rather than to the map alone.

\subsection{Dynamic-tune benchmark}
\label{subsec:pure_runs}
The values in Table~\ref{tab:pure_runs} are computed after the common $30$-frame tracker warm-up exclusion, and the bottom row reports the unweighted geometric mean across the $15$ cells, $\mathrm{gm} = \bigl(\prod_{i=1}^{15} \mathrm{MAE}_i\bigr)^{1/15}$, which down-weights outlier cells relative to the arithmetic mean.

\begin{table}[htbp]
\centering
\caption{Main dynamic-tune mean absolute error in $10^{-3}$ tune units across the $15$ (SNR, shape) combinations on broadband-noise spectra. Bold marks the smallest MAE per row, and the smallest geometric mean on the bottom row. \label{tab:pure_runs}}
\smallskip
\footnotesize
\setlength{\tabcolsep}{4pt}
\begin{tabular}{c c c c c c c c c}
\toprule
SNR (dB) & shape & deep-learning & classical & CNN+KF & T-PD & PD & CF & WC \\
\midrule
\multirow{3}{*}{$-20$} & slow   & \textbf{0.456} & 3.375 & 1.031 & 0.638 & 60.960 & 61.827 & 62.316 \\
                       & medium & \textbf{0.475} & 5.200 & 1.565 & 1.041 & 60.398 & 61.173 & 61.698 \\
                       & fast   & \textbf{0.892} & 10.612 & 2.981 & 2.922 & 59.420 & 60.124 & 60.701 \\
\midrule
\multirow{3}{*}{$-15$} & slow   & \textbf{0.209} & 0.326 & 0.300 & 0.407 & 19.634 & 19.219 & 20.832 \\
                       & medium & \textbf{0.210} & 0.477 & 0.412 & 0.749 & 19.183 & 18.876 & 20.324 \\
                       & fast   & \textbf{0.345} & 0.916 & 0.783 & 1.847 & 19.094 & 18.733 & 20.309 \\
\midrule
\multirow{3}{*}{$-10$} & slow   & \textbf{0.140} & 0.158 & 0.164 & 0.342 & 2.492 & 1.839 & 2.439 \\
                       & medium & \textbf{0.140} & 0.193 & 0.192 & 0.740 & 2.933 & 2.238 & 2.857 \\
                       & fast   & \textbf{0.189} & 0.355 & 0.295 & 1.957 & 2.949 & 2.269 & 2.877 \\
\midrule
\multirow{3}{*}{$-5$}  & slow   & 0.102 & 0.098 & \textbf{0.086} & 0.326 & 0.896 & 0.278 & 0.361 \\
                       & medium & 0.100 & 0.123 & \textbf{0.090} & 0.739 & 0.885 & 0.238 & 0.332 \\
                       & fast   & 0.131 & 0.261 & \textbf{0.113} & 2.025 & 0.876 & 0.244 & 0.343 \\
\midrule
\multirow{3}{*}{$0$}   & slow   & 0.085 & 0.087 & \textbf{0.053} & 0.326 & 0.793 & 0.187 & 0.154 \\
                       & medium & 0.083 & 0.108 & \textbf{0.051} & 0.743 & 0.795 & 0.170 & 0.141 \\
                       & fast   & 0.116 & 0.245 & \textbf{0.058} & 2.033 & 0.794 & 0.176 & 0.148 \\
\midrule
\multicolumn{2}{l}{geometric mean} & \textbf{0.187} & 0.418 & 0.243 & 0.873 & 4.693 & 2.557 & 2.809 \\
\bottomrule
\end{tabular}
\end{table}

\begin{figure}[htbp]
\centering
\includegraphics[width=\linewidth]{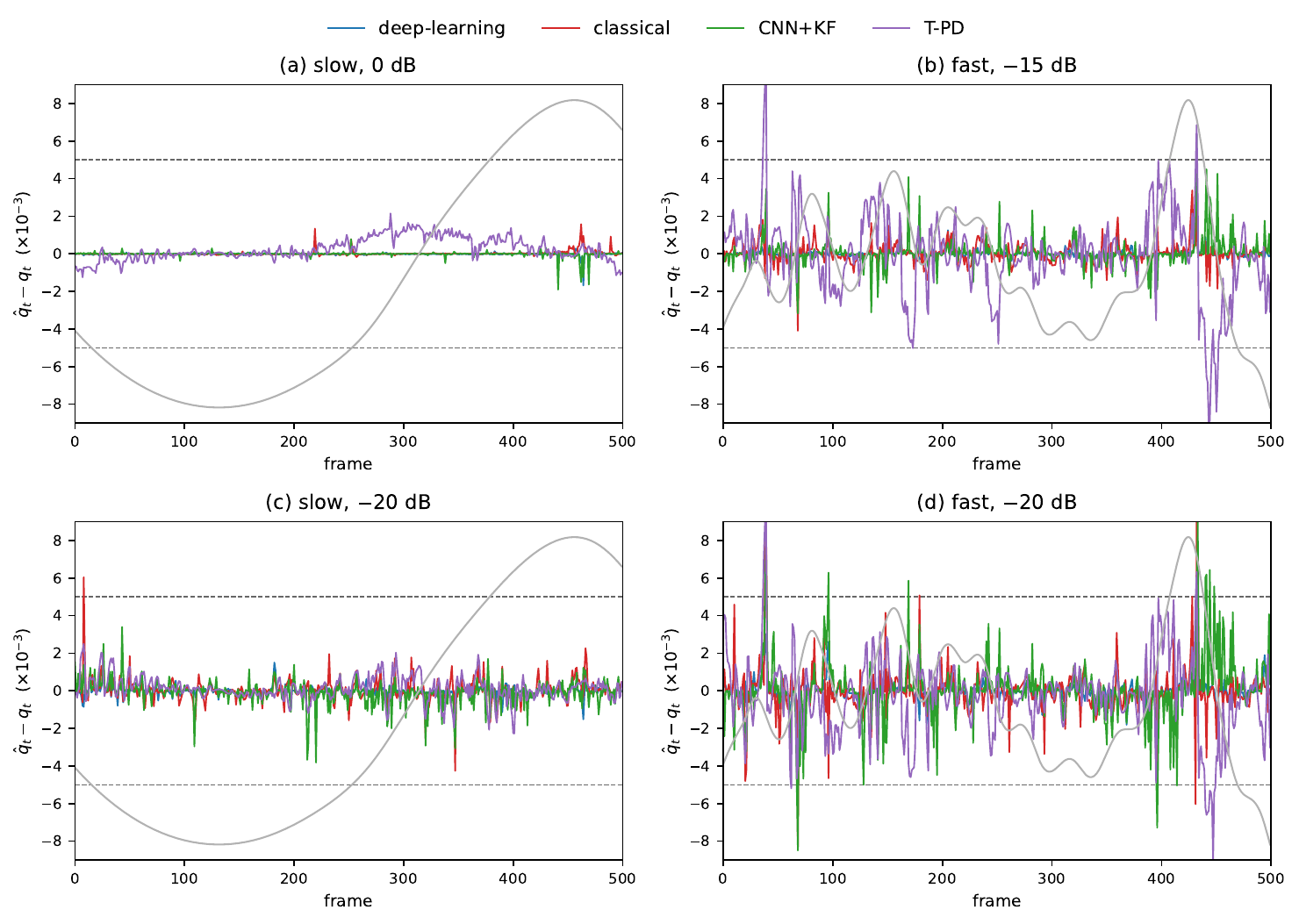}
\caption{Per-frame signed tune error $\hat q_t - q_t$ (in $10^{-3}$ tune units) of the four temporal estimators on a $500$-frame post-warm-up slice of four representative cells of the dynamic-tune benchmark: (a)~the high-SNR slow trajectory ($0$\,dB) where all estimators agree, (b)~the fast trajectory at $-15$\,dB where the latency-compensated T-PD's wide-EMA pool lags the moving sideband, (c)~the slow trajectory at the deepest SNR, and (d)~the fast trajectory at the deepest SNR where the classical estimator's excursions are largest. The faint grey curve in each panel is the ground-truth tune trajectory (right axis, arbitrary scale); the dashed lines mark the $\pm 5 \times 10^{-3}$ catastrophic-miss threshold. T-PD is shown after its latency compensation ($N^{*} = 3$; Section~\ref{subsec:setup}). Vertical axes are clipped to $\pm 9 \times 10^{-3}$. \label{fig:dynamic_traj}}
\end{figure}

\paragraph{Operating-regime split.}
The deep-learning estimator holds the aggregate geometric-mean advantage, and the cell-by-cell ordering splits along a clean SNR boundary. It is best in $9$ of the $15$ cells (every trajectory shape at $-20$, $-15$, and $-10$\,dB), where its likelihood-map output retains the multi-modal spectral evidence that the discrete $(q, v)$ filter integrates across frames; its margin over every alternative widens toward the deep-SNR corner, where it is the only estimator that keeps every shape below the $10^{-3}$ target. The CNN+KF baseline takes the six easy cells at $-5$ and $0$\,dB, consistent with its stronger single-frame readout in the static comparison of Section~\ref{subsec:static}; but there every deployable temporal estimator already sits far below the catastrophic-miss threshold, so that ranking carries little operational weight. The classical estimator and the latency-compensated T-PD win no cell: the classical estimator's motion-compensated matched-filter pool holds the mid-SNR band but degrades steeply at the deepest SNR, and T-PD's wide-EMA pool smears the moving sideband on the faster trajectories at every SNR. The accuracy case for the deep-learning estimator therefore concentrates exactly where accuracy is scarce, at and below $-10$\,dB, while the high-SNR cells discriminate little in operational terms; this pattern informs the deployment guidance in Section~\ref{subsec:decision_matrix}.

\paragraph{Catastrophic-miss rate.}
The mean absolute tune error of Table~\ref{tab:pure_runs} averages over all post-warm-up frames and so masks a tail-risk failure mode that is operationally important: an estimator can have a competitive average MAE while suffering a non-trivial fraction of frames whose individual error exceeds a threshold. We therefore report the catastrophic-miss rate $T_{5 \cdot 10^{-3}} = \mathrm{P}(\lvert\hat q_t - q_t\rvert > 5 \times 10^{-3})$ as a tail-risk metric, summarised across the $15$ cells in Table~\ref{tab:pure_runs_tail}.

\begin{table}[htbp]
\centering
\caption{Catastrophic-miss rate $T_{5 \cdot 10^{-3}}$, per-estimator summary across the $15$ cells of Table~\ref{tab:pure_runs}. ``mean'' and ``max'' are the arithmetic mean and maximum of the per-cell catastrophic-miss fraction, expressed as percentages of post-warm-up frames. ``cells $> 1\%$'' and ``cells $> 10\%$'' count the cells in which $T_{5 \cdot 10^{-3}}$ exceeds the indicated threshold. Bold marks the smallest value per row. \label{tab:pure_runs_tail}}
\smallskip
\footnotesize
\begin{tabular}{l c c c c c c c}
\toprule
metric & deep-learning & classical & CNN+KF & T-PD & PD & CF & WC \\
\midrule
mean $T_{5 \cdot 10^{-3}}$ (\%) & \textbf{0.14} & 1.75 & 1.82 & 3.57 & 15.24 & 12.94 & 19.50 \\
max $T_{5 \cdot 10^{-3}}$ (\%)  & \textbf{1.77} & 13.35 & 17.83 & 13.60 & 46.14 & 47.77 & 66.49 \\
cells $> 1\%$ & \textbf{1} & 4 & 4 & 6 & 15 & 9 & 9 \\
cells $> 10\%$ & \textbf{0} & 1 & 1 & 3 & 6 & 6 & 6 \\
\bottomrule
\end{tabular}
\end{table}

\paragraph{Tail-risk separation.}
On the catastrophic-miss metric the deep-learning estimator is the clear outlier in the safe direction: it is the only estimator that never crosses the $10\%$ threshold in any cell, and it crosses the $1\%$ threshold in just one (the deepest-SNR fast corner), whereas every other temporal estimator crosses it in several cells and the single-frame baselines exceed $10\%$ repeatedly. The separation is driven almost entirely by the $-20$\,dB row, where the per-frame spectral evidence is weakest. The deep-learning estimator's per-frame threshold-exceedance risk is therefore an order of magnitude below every alternative, a reliability difference that the average MAE alone does not capture.

\subsection{Robustness studies}
\label{subsec:robustness}
The main dynamic-tune benchmark of Section~\ref{subsec:pure_runs} characterises the two proposed estimators within the training envelope. We now group three stress tests that probe their behaviour at or beyond the edges of that envelope: recovery of tracking after a beam-off interval between acceleration cycles (Section~\ref{subsec:beam_loss}), tolerance to residual narrow-band interference (Section~\ref{subsec:narrowband}), and degradation at per-frame signal-to-noise ratios beyond both edges of the training range (Section~\ref{subsec:ood_snr}).

\subsubsection{Robustness to beam loss}
\label{subsec:beam_loss}
The dynamic-tune sequences of Section~\ref{subsec:pure_runs} contain only smoothly varying trajectories, which exercise tracking under continuous tune motion but not the recovery of lock after an abrupt change in the observed tune. Within a single energy ramp the tune varies continuously, without abrupt jumps. The abrupt change of operational interest arises instead from the cyclic operation of a compact proton-therapy synchrotron, which repeats an injection--acceleration--slow-extraction sequence: the beam is removed at the end of each cycle and a fresh beam is injected at the start of the next, so the tune observed when a beam-bearing interval resumes can sit at a working point displaced from the last tracked state by far more than a single per-frame step. A temporally regularised estimator must then re-lock onto the new tune rather than continue following the stale prior, and one that takes many frames to relock would misreport the tune over an operationally significant fraction of the short acceleration window. We reproduce this scenario directly: each test sequence follows a beam-present segment of the dynamic-tune stream, interposes a beam-off interval of pure detector noise with no betatron sideband to model the inter-cycle gap, and resumes with a beam-present segment whose tune is displaced by an injected step $\Delta q$; the two proposed estimators are reported side by side with the CNN+KF baseline of~\cite{sun2026real} and the latency-compensated T-PD baseline of~\cite{sun2025high} as reference temporally regularised pipelines.

\paragraph{Stress-test construction.}
We build $350$-frame recovery sequences from the slow-trajectory streams of the dynamic-tune benchmark of Section~\ref{subsec:pure_runs} (the same varying-acquisition-parameter source and preprocessing). Each sequence concatenates a $150$-frame beam-present pre-segment taken from the source stream, a $50$-frame beam-off gap of pure detector noise with no betatron sideband (white noise matched to the per-frame noise floor of the last pre-gap frame), and a $150$-frame beam-present post-segment drawn from a \emph{second} interval of the same stream whose working point is displaced from the pre-gap value by the displacement $\Delta q$ (a small constant tune-bin shift, below half a per-frame step on average, sets the displacement exactly). We sweep $\Delta q \in \{5, 25, 100, 400\} \times 10^{-3}$ at the three SNR levels $\{-20, -15, -10\}$\,dB where recovery is non-trivial (at $-5$\,dB and above every temporal estimator re-locks within about a frame, so those cells discriminate nothing); for each $(\mathrm{SNR}, \Delta q)$ cell we draw $10$ sequences at spread start indices, giving $120$ sequences in total. Each estimator runs continuously across the three segments and is not told that the beam is absent during the gap, so its temporal state must survive the beam-off interval and re-lock onto the displaced tune once the beam returns; the post-gap behaviour is therefore attributable entirely to each estimator's temporal regularisation. The three stateless single-frame baselines of Section~\ref{subsec:pure_runs} (PD, CF, WC) are omitted because, lacking any temporal state, they cannot exhibit recovery dynamics; their per-frame behaviour on beam-present frames is already reported in Table~\ref{tab:pure_runs}.

\begin{table}[htbp]
\centering
\caption{Beam-loss recovery across a beam-off gap (construction described in the text), with the displacement $\Delta q$ in $10^{-3}$ tune units. The reported metric is the recovery frame count $N_{\mathrm{rec}}$: the number of post-gap frames until the per-frame error first stays below the catastrophic-miss threshold of $5 \times 10^{-3}$ (Section~\ref{subsec:pure_runs}) for $W = 10$ consecutive frames, i.e.\ until the estimator stops catastrophically missing; a re-lock criterion at the $10^{-3}$ accuracy target would instead be unsatisfiable in steady state for the baseline estimators in the $-20$\,dB cells and would conflate re-acquisition speed with steady-state accuracy. The count is non-integer because it is the mean over the $10$ sequences of an integer per-sequence count. A parenthesised count marks sequences (out of $10$) that never re-lock within the post-gap segment; $N_{\mathrm{rec}}$ averages the recovering sequences. The classical, CNN+KF, and T-PD estimators recover on all $120$ of their sequences; the deep-learning estimator fails to re-lock on one sequence (at $-20$\,dB and $\Delta q = 400 \times 10^{-3}$). Bold marks the smallest $N_{\mathrm{rec}}$ per row. \label{tab:beam_loss}}
\smallskip
\footnotesize
\setlength{\tabcolsep}{4pt}
\begin{tabular}{c c c c c c}
\toprule
SNR & $\Delta q$ & \multicolumn{4}{c}{recovery frame count $N_{\mathrm{rec}}$} \\
\cmidrule(lr){3-6}
(dB) & ($10^{-3}$) & deep-learning & classical & CNN+KF & T-PD \\
\midrule
\multirow{4}{*}{$-20$} & $5$   & 2.6 & 1.2 & \textbf{0.6} & \textbf{0.6} \\
                       & $25$  & 1.5 & 0.3 & \textbf{0.2} & 3.9 \\
                       & $100$ & 1.6 & 0.8 & \textbf{0.1} & 3.3 \\
                       & $400$ & 2.3 (1) & 2.2 & \textbf{0.3} & 20.1 \\
\midrule
\multirow{4}{*}{$-15$} & $5$   & 0.3 & 0.6 & \textbf{0.0} & 0.6 \\
                       & $25$  & 0.5 & 0.1 & \textbf{0.0} & 0.6 \\
                       & $100$ & 0.4 & \textbf{0.1} & \textbf{0.1} & 1.9 \\
                       & $400$ & 1.3 & \textbf{0.0} & \textbf{0.0} & 3.4 \\
\midrule
\multirow{4}{*}{$-10$} & $5$   & 0.4 & \textbf{0.0} & \textbf{0.0} & 0.4 \\
                       & $25$  & 0.6 & \textbf{0.0} & \textbf{0.0} & 0.6 \\
                       & $100$ & 0.7 & \textbf{0.0} & \textbf{0.0} & 0.6 \\
                       & $400$ & 1.2 & \textbf{0.4} & 0.9 & 1.4 \\
\bottomrule
\end{tabular}
\end{table}

\begin{figure}[htbp]
\centering
\includegraphics[width=\linewidth]{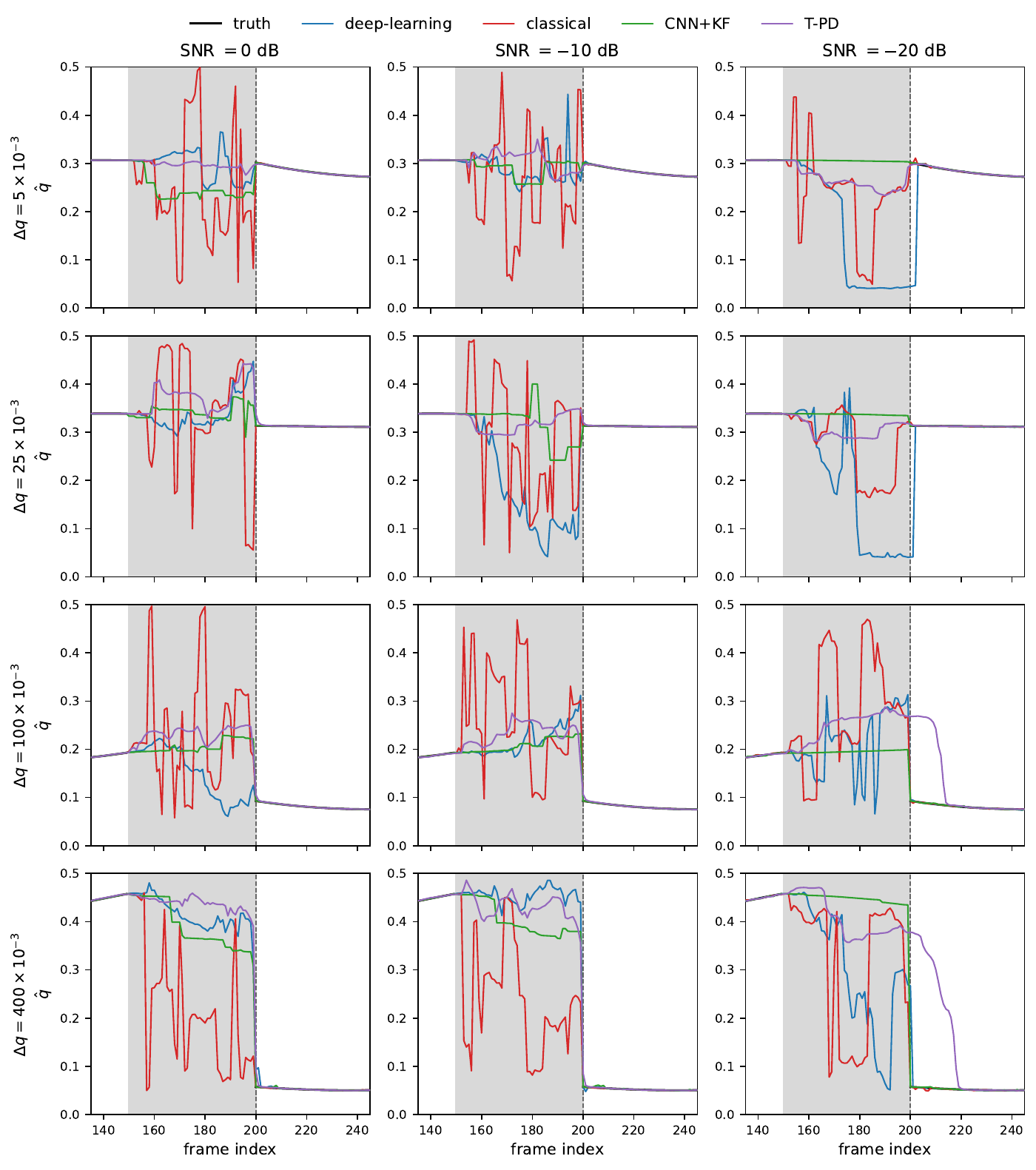}
\caption{Representative per-frame recovery trajectories across the beam-loss sweep: rows are the displacement magnitude $\Delta q \in \{5, 25, 100, 400\} \times 10^{-3}$, columns three representative SNR levels ($0$, $-10$, and $-20$\,dB). Each panel shows the ground-truth tune (black) and the four temporal estimators on one representative sequence, with the beam-off gap shaded and the dashed line marking beam return. The panels illustrate the recovery \emph{mechanisms} (the shape and size of each estimator's post-return excursion and its behaviour on the beam-off noise), not the $N_{\mathrm{rec}}$ ordering of Table~\ref{tab:beam_loss}: the re-lock criterion ($5 \times 10^{-3}$ held for $10$ frames) is one per cent of the plotted axis and is resolved in the table, whose entries additionally average ten sequences per cell, including slow and failed recoveries that a single displayed sequence does not represent. T-PD is shown after the $N^{*} = 3$ latency compensation (Section~\ref{subsec:setup}); vertical axes span the observable $[0, 0.5)$ band. \label{fig:recovery_traj}}
\end{figure}

\paragraph{Recovery reliability.}
Recovery is near-universal on this sweep: of the $480$ estimator-sequences ($4$ estimators $\times$ $120$), only one fails to re-lock within the post-gap segment, a single deep-learning-estimator sequence at $-20$\,dB and the widest displacement; the classical, CNN+KF, and T-PD estimators recover on every one of their $120$ sequences. Where recovery succeeds, every estimator's post-recovery tail accuracy returns to its dynamic-tune level of Table~\ref{tab:pure_runs}, so the beam-off disturbance leaves no lasting bias once lock is regained; the single non-recovering sequence instead settles far from the true working point and is excluded from that comparison. Table~\ref{tab:beam_loss} reports the recovery frame count $N_{\mathrm{rec}}$, and Figure~\ref{fig:recovery_traj} shows representative per-frame trajectories, including the post-return excursion each estimator passes through before re-locking.

\paragraph{Recovery speed.}
All four estimators re-lock quickly on this near-stationary baseline: the deep-learning, classical, and CNN+KF estimators re-lock within three frames in every cell, so the beam-loss axis no longer separates them as sharply as the deep-SNR accuracy of Table~\ref{tab:pure_runs} does. The residual ordering favours the estimators with the least temporal memory to unwind. The per-frame CNN+KF baseline re-locks fastest overall, within a single frame in every cell: its scalar Kalman filter runs at near-unit gain on the confident per-frame CNN regression, so a single clean returned-beam frame overrides the stale filter state with little prior to unwind; the classical estimator's shallow two-frame motion-compensated pool follows close behind. The deep-learning estimator is a frame or two slower: its $(q, v)$ posterior diffuses toward uniform during the beam-off gap and must re-concentrate on the displaced working point once the beam returns, a re-acquisition that costs a small handful of frames and, on one $-20$\,dB widest-displacement sequence, fails to complete within the post-gap segment because the velocity-bounded tracker cannot slew across the widest displacement fast enough for the weak returned-beam map to guide it back. The latency-compensated T-PD baseline is slowest, its wide temporal pool slow to flush the stale pre-gap tune (up to $20$ frames at the $-20$\,dB widest displacement). The very temporal integration that gives the deep-learning estimator its deep-low-SNR accuracy advantage (Section~\ref{subsec:pure_runs}) is therefore a mild liability for abrupt re-lock: on this axis the shorter-memory readouts recover marginally faster. Because all four estimators re-lock within a few frames of the millisecond-scale acquisition window, the beam-loss axis is a weak deployment discriminator rather than a decisive one.

\subsubsection{Robustness to residual narrow-band interference}
\label{subsec:narrowband}
The main dynamic-tune benchmark of Section~\ref{subsec:pure_runs} characterises the two proposed estimators on broadband-noise spectra. We now test robustness to a residual narrow-band line that survives the upstream reference-spectrum subtraction and per-frequency electrode decoupling described in Section~\ref{subsec:real_beam}. Such a line sits at a fixed \emph{absolute} frequency (a digital clock, switching-supply, or RF-system harmonic), so how it appears on the folded-tune grid is set by the machine operating regime through the revolution frequency $f_{\mathrm{rev}}$ that defines the tune normalisation. Two regimes bracket deployment. During slow extraction the beam sits on a flat-top and $f_{\mathrm{rev}}$ is constant, so a fixed absolute line maps to a \emph{stationary} tune position; during acceleration $f_{\mathrm{rev}}$ ramps, so the same absolute line \emph{drifts} across the tune grid. An estimator that stays accurate under both is the more robust candidate for a real, spectrally complex environment.

\paragraph{Slow extraction (stationary clutter).}
The first regime uses a single flat-top stream with the revolution frequency held constant at $6$\,MHz and the betatron tune fixed at the third-integer slow-extraction operating point ($q \approx 0.333$ with a $\pm 0.001$ jitter, about $0.03$ bins per frame, so the beam is essentially stationary), so a fixed absolute-frequency line maps to a fixed tune bin. The stream is $600$ frames (about $0.6$\,s at the $1$\,ms cadence), evaluated at the same five SNR levels as the acceleration regime below. Three Gaussian peaks are injected at fixed bin positions and held across frames while their amplitudes are resampled per frame in the $1$ to $2\times$ betatron-sideband range, a subset of the training-augmentation range that models a post-suppression residual. All estimators are reported under their deployed configuration, after the common $30$-frame tracker warm-up exclusion.

\paragraph{Acceleration (drifting clutter).}
The second regime evaluates the same contaminant on a single representative energy ramp ($T = 600$ frames, about $0.6$\,s at the $1$\,ms frame cadence, with $f_{\mathrm{rev}}$ sweeping $1.5 \to 7.5$\,MHz). Three lines are placed at fixed absolute frequencies in the $34$--$42$\,MHz band (within the acquisition band of the ramp stream) and injected before the tune mapping, so their folded-tune positions drift as $f_{\mathrm{rev}}$ ramps; per-frame amplitudes use the same $1$ to $2\times$ calibration, verified to reproduce the target mapped amplitude. A ramp is a single continuous trajectory whose natural unit is its duration rather than a count of independent draws; the figures below are a geometric mean over the five SNR levels of this one ramp, with the per-frame statistics estimated over the post-warm-up frames, all but eight of which are free of a sideband transit. Those eight transit frames (nine over the full ramp, one of which falls in the warm-up window) are too few to support a separate crossing-specific number (a consequence of the ramp rate, at which a fixed absolute line sweeps past the sideband within a frame), and are not reported as one.

\begin{table}[htbp]
\centering
\caption{Robustness to a residual narrow-band line across the two machine operating regimes: mean absolute tune error in $10^{-3}$ tune units. \emph{Extraction}: on the flat-top ($f_{\mathrm{rev}}$ constant at $6$\,MHz) a fixed-absolute-frequency line maps to a fixed (stationary) tune bin, over a single $0.6$\,s ($600$-frame) flat-top stream with the tune held at the slow-extraction operating point $q \approx 0.333$, at five SNR levels. \emph{Acceleration}: the same line ($f_{\mathrm{rev}}$ ramping $1.5 \to 7.5$\,MHz) drifts across the tune grid, over a single $0.6$\,s ($600$-frame) ramp at five SNR levels. Both regimes report the deployed configuration after the $30$-frame warm-up exclusion; T-PD uses $N^\ast = 3$ latency compensation. Bold marks the smallest MAE per row. Column abbreviations as in Table~\ref{tab:pure_runs}. \label{tab:narrowband}}
\smallskip
\footnotesize
\setlength{\tabcolsep}{4pt}
\begin{tabular}{l c c c c c c c c c}
\toprule
regime & SNR (dB) & shape & deep-learning & classical & CNN+KF & T-PD & PD & CF & WC \\
\midrule
\multirow{5}{*}{\rotatebox[origin=c]{90}{extraction}}
 & $-20$ & flat & \textbf{0.121} & 0.166 & 0.122 & 0.130 & 0.189 & 0.338 & 1.978 \\
 & $-15$ & flat & \textbf{0.040} & 0.067 & 0.041 & 0.123 & 0.141 & 0.105 & 0.556 \\
 & $-10$ & flat & 0.017 & 0.026 & \textbf{0.014} & 0.123 & 0.134 & 0.031 & 0.150 \\
 & $-5$  & flat & 0.013 & 0.011 & \textbf{0.006} & 0.121 & 0.133 & 0.034 & 4.811 \\
 & $0$   & flat & 0.016 & 0.005 & \textbf{0.003} & 0.122 & 0.133 & 0.027 & 5.975 \\
\cmidrule(lr){2-10}
\multicolumn{3}{l}{\hspace{1em}$5$-cell extraction geometric mean} & 0.028 & 0.028 & \textbf{0.017} & 0.124 & 0.145 & 0.063 & 1.366 \\
\midrule
\multirow{5}{*}{\rotatebox[origin=c]{90}{acceleration}}
 & $-20$ & ramp & \textbf{0.529} & 1.229 & 6.972 & 1.146 & 44.056 & 43.861 & 46.317 \\
 & $-15$ & ramp & \textbf{0.192} & 0.448 & 0.370 & 0.760 & 6.192 & 5.431 & 7.958 \\
 & $-10$ & ramp & \textbf{0.074} & 0.240 & 0.094 & 0.575 & 4.826 & 4.320 & 5.999 \\
 & $-5$  & ramp & 0.051 & 0.198 & \textbf{0.048} & 0.562 & 4.696 & 4.329 & 5.756 \\
 & $0$   & ramp & \textbf{0.044} & 0.190 & 0.081 & 0.556 & 4.647 & 4.321 & 5.688 \\
\cmidrule(lr){2-10}
\multicolumn{3}{l}{\hspace{1em}$5$-cell acceleration geometric mean} & \textbf{0.111} & 0.346 & 0.248 & 0.690 & 7.792 & 7.193 & 9.374 \\
\bottomrule
\end{tabular}
\end{table}

\paragraph{Results and clutter rejection.}
In the extraction regime every temporal estimator absorbs the stationary line with little change from its clean flat-top reference, and the regime discriminates weakly among them: all four trackers stay far below any operational threshold, so the stationary-clutter case is effectively solved by every one of them. The single-frame baselines split with the readout, the peak-detection and curve-fitting readouts staying near the broadband-noise floor because a stationary strong sideband still wins most per-frame argmaxes, while the windowed-centroid readout is repeatedly captured by the line. Holding the tune constant places both the sideband and the line at $v \approx 0$, so the velocity prior carries no discriminating signal here: the rejection of the stationary clutter is attributable to the spectral discrimination of the CNN maps and, for the classical estimator, to the matched filter's frequency selectivity. The velocity prior is the additional mechanism that handles the drifting clutter of the acceleration regime.

Under acceleration the contrast sharpens into the section's main result, and the ordering is set by how much temporal context each estimator carries. The velocity-aware deep-learning posterior is essentially immune to the drifting line: it treats the line as motion inconsistent with the beam, holds its clean-ramp accuracy, wins almost every cell, and never catastrophically misses. The wide-pool T-PD baseline matches that immunity by a different route, averaging the brief transits away over its long temporal window. The remaining estimators are progressively drawn off the beam as their temporal memory shrinks: the classical estimator's short motion-compensated pool is pulled toward the line only on the frames where it is strong; the per-frame CNN+KF regression is dragged onto the line often enough to raise a heavy deep-SNR tail, its Kalman filter following the strongest peak rather than the beam; and the memoryless single-frame readouts are captured outright, their error set by the line's excursion across the grid. This drifting-clutter regime is the one that actually stresses clutter rejection, and it is where the velocity prior is most clearly load-bearing; under the stationary clutter of the extraction regime, where every estimator sees the line at one fixed position, it is instead one of several trackers that all absorb it.

\paragraph{Catastrophic-miss rate.}
We report the catastrophic-miss rate $T_{5 \cdot 10^{-3}}$ (defined in Section~\ref{subsec:pure_runs}) across the narrow-band cells in Table~\ref{tab:narrowband_tail}.

\begin{table}[htbp]
\centering
\caption{Catastrophic-miss rate $T_{5 \cdot 10^{-3}}$ under the residual narrow-band line, per-estimator summary. ``mean'' and ``max'' are the arithmetic mean and maximum of the per-cell catastrophic-miss fraction (percent of post-warm-up frames) over the $5$ extraction cells and the $5$ acceleration cells of Table~\ref{tab:narrowband}. Bold marks the smallest value per row. Column abbreviations as in Table~\ref{tab:pure_runs}. \label{tab:narrowband_tail}}
\smallskip
\footnotesize
\begin{tabular}{l l c c c c c c c}
\toprule
regime & metric & deep-learning & classical & CNN+KF & T-PD & PD & CF & WC \\
\midrule
\multirow{2}{*}{extraction} & mean (\%) & \textbf{0.00} & \textbf{0.00} & \textbf{0.00} & \textbf{0.00} & \textbf{0.00} & \textbf{0.00} & 26.39 \\
 & max (\%) & \textbf{0.00} & \textbf{0.00} & \textbf{0.00} & \textbf{0.00} & \textbf{0.00} & \textbf{0.00} & 79.12 \\
\midrule
\multirow{2}{*}{acceleration} & mean (\%) & \textbf{0.00} & 1.12 & 2.46 & 0.63 & 12.98 & 8.91 & 25.51 \\
 & max (\%) & \textbf{0.00} & 5.26 & 11.75 & 3.00 & 37.02 & 31.75 & 62.81 \\
\bottomrule
\end{tabular}
\end{table}

Under stationary extraction clutter every temporal estimator keeps tail risk at zero, and among the baselines only the windowed-centroid readout shows a double-digit maximum. Under acceleration the tail metric sharpens the same ordering: the deep-learning estimator alone is entirely free of catastrophic misses, the wide-pool T-PD stays in the low single digits, and the remaining estimators' catastrophic-miss rates climb as their temporal memory shrinks, reaching double digits for the per-frame CNN+KF baseline and the stateless readouts. Only the velocity-aware estimator holds tail risk at zero under the drifting line.

\subsubsection{Robustness to out-of-distribution SNR}
\label{subsec:ood_snr}
The deep-learning estimator is trained on the $[-20, 0]$\,dB SNR envelope of Section~\ref{subsec:setup}, on which both proposed estimators are configured, and the published baselines T-PD~\cite{sun2025high} and CNN+KF~\cite{sun2026real} are designed and validated only for SNR not below $-20$\,dB. We probe both edges of this envelope on clean broadband-noise spectra (above the upper edge and below the common lower edge) to verify that the proposed estimators do not regress catastrophically outside their training support and to characterise how each estimator degrades there.

We first extend the main dynamic-tune benchmark above the upper edge. For each SNR $\in \{+5, +10, +15, +20\}$\,dB and each trajectory shape $\in \{\text{slow}, \text{medium}, \text{fast}\}$, a $10\,000$-frame dynamic-tune stream is generated with the same simulator and grid as Section~\ref{subsec:setup}, and all seven estimators are run on the bit-identical input stream after the common $30$-frame warm-up exclusion. The T-PD column reports the latency-compensated MAE with the same $N^{*}$ used in Section~\ref{subsec:pure_runs}.

\begin{table}[htbp]
\centering
\caption{Out-of-distribution high-SNR evaluation: per-cell MAE in $10^{-3}$ tune units on broadband-noise spectra at $+5$, $+10$, $+15$, and $+20$\,dB (training spans $-20$ to $0$\,dB). Bold marks the smallest MAE per row across the seven estimators. \label{tab:highsnr}}
\smallskip
\footnotesize
\setlength{\tabcolsep}{4pt}
\begin{tabular}{c c c c c c c c c}
\toprule
SNR (dB) & shape & deep-learning & classical & CNN+KF & T-PD & PD & CF & WC \\
\midrule
\multirow{3}{*}{$+5$}  & slow   & 0.087 & 0.086 & \textbf{0.052} & 0.323 & 0.781 & 0.172 & 0.113 \\
                       & medium & 0.084 & 0.107 & \textbf{0.052} & 0.753 & 0.784 & 0.158 & 0.107 \\
                       & fast   & 0.116 & 0.245 & \textbf{0.057} & 2.031 & 0.785 & 0.161 & 0.112 \\
\midrule
\multirow{3}{*}{$+10$} & slow   & 0.091 & 0.086 & \textbf{0.066} & 0.324 & 0.780 & 0.169 & 0.107 \\
                       & medium & 0.088 & 0.107 & \textbf{0.064} & 0.752 & 0.784 & 0.156 & 0.102 \\
                       & fast   & 0.120 & 0.245 & \textbf{0.072} & 2.036 & 0.780 & 0.160 & 0.107 \\
\midrule
\multirow{3}{*}{$+15$} & slow   & 0.093 & 0.086 & \textbf{0.071} & 0.323 & 0.780 & 0.170 & 0.106 \\
                       & medium & 0.089 & 0.107 & \textbf{0.070} & 0.753 & 0.783 & 0.156 & 0.100 \\
                       & fast   & 0.121 & 0.246 & \textbf{0.079} & 2.035 & 0.781 & 0.160 & 0.107 \\
\midrule
\multirow{3}{*}{$+20$} & slow   & 0.093 & 0.086 & \textbf{0.073} & 0.323 & 0.780 & 0.169 & 0.106 \\
                       & medium & 0.089 & 0.107 & \textbf{0.071} & 0.753 & 0.783 & 0.156 & 0.101 \\
                       & fast   & 0.122 & 0.246 & \textbf{0.080} & 2.036 & 0.781 & 0.160 & 0.106 \\
\midrule
\multicolumn{2}{l}{geometric mean} & 0.098 & 0.131 & \textbf{0.067} & 0.791 & 0.782 & 0.162 & 0.106 \\
\bottomrule
\end{tabular}
\end{table}

At the opposite edge we drop five decibels below the common lower edge of the training and published-baseline design ranges. For each trajectory shape a $10\,000$-frame clean dynamic-tune stream is generated at a nominal SNR of $-25$\,dB using the simulator and grid of Section~\ref{subsec:setup}, and all seven estimators are run on bit-identical input streams under their deployed configuration after the common $30$-frame warm-up exclusion.

\begin{table}[htbp]
\centering
\caption{Extreme-low-SNR stress test: per-cell MAE in $10^{-3}$ tune units on clean spectra at nominal SNR $= -25$\,dB ($5$\,dB below the lower edge of the training and published-baseline design envelopes), for the three trajectory shapes. Bold marks the smallest MAE per row across the seven estimators. Column abbreviations as in Table~\ref{tab:pure_runs}. \label{tab:lowsnr_stress}}
\smallskip
\footnotesize
\setlength{\tabcolsep}{4pt}
\begin{tabular}{l c c c c c c c}
\toprule
shape & deep-learning & classical & CNN+KF & T-PD & PD & CF & WC \\
\midrule
slow           & \textbf{3.638} & 96.075 & 65.833 & 22.746 & 120.940 & 122.158 & 119.856 \\
medium         & \textbf{4.379} & 97.070 & 66.495 & 27.432 & 120.130 & 121.277 & 118.998 \\
fast           & \textbf{6.341} & 95.882 & 64.245 & 58.167 & 117.284 & 118.397 & 116.003 \\
\midrule
geometric mean & \textbf{4.657} & 96.341 & 65.518 & 33.109 & 119.441 & 120.600 & 118.274 \\
\bottomrule
\end{tabular}
\end{table}

\begin{figure}[htbp]
\centering
\includegraphics[width=\linewidth]{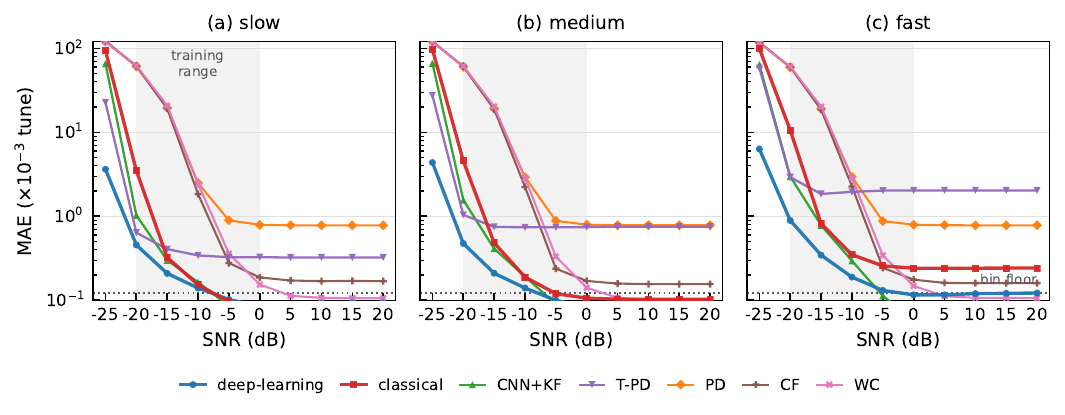}
\caption{Out-of-distribution behaviour across the full SNR span: per-cell mean absolute tune error ($10^{-3}$ tune units, logarithmic vertical axis) on clean broadband-noise spectra versus SNR, for the (a)~slow, (b)~medium and (c)~fast trajectory shapes; one curve per estimator. The ten SNR points are assembled from Tables~\ref{tab:pure_runs}, \ref{tab:highsnr} and~\ref{tab:lowsnr_stress}, which share the same clean-broadband simulator and grid so the axis is a single continuous condition. The shaded band marks the $-20$ to $0$\,dB training range and the dotted line the bin-quantisation reference level $\Delta_{q}/4 \approx 0.122 \times 10^{-3}$ tune units; T-PD is shown after its latency compensation ($N^{*} = 3$; Section~\ref{subsec:setup}). \label{fig:ood_snr}}
\end{figure}

\paragraph{Bin-resolution floor and high-SNR behaviour.}
At sufficiently high SNR the broadband-noise contribution to the per-frame readout falls to the scale of the discretisation of the $L = 1024$-bin folded-tune grid; the bin-quantisation reference level is $\Delta_{q} / 4 \approx 1.22 \times 10^{-4}$ tune units (Section~\ref{subsec:pure_runs}). Figure~\ref{fig:ood_snr} traces each estimator's per-cell MAE across the full $-25$ to $+20$\,dB span against this reference, combining the in-distribution benchmark of Table~\ref{tab:pure_runs} with the high-SNR cells of Table~\ref{tab:highsnr} and the extreme-low-SNR cells of Table~\ref{tab:lowsnr_stress}. At high SNR the CNN+KF baseline is the most accurate estimator in every cell and holds the geometric mean: its scalar regression head feeds a per-frame uncertainty channel that the Kalman tracker rescales into measurement variance, and on clean spectra this calibrated-gain smoothing converges to a sub-bin readout well below the quantisation reference. The deep-learning estimator sits second, also at or below the reference on the slow and medium trajectories; its residual is set by the finite width the discrete posterior retains (about $0.6$ bins, Section~\ref{subsec:ablation}) rather than by the map, and it neither collapses nor overtakes: the $+5$ to $+20$\,dB cells reproduce the $-5$/$0$\,dB ordering of Table~\ref{tab:pure_runs}, with saturation in place of further gain. The classical estimator matches the leaders on the slow cells but is lag-limited on the fast ones, where the residual lag of its two-frame motion-compensated pool dominates once the noise floor is gone; the stateless WC and CF readouts settle between the CNN pair and the classical fast cells, and the latency-compensated T-PD remains several times above the leaders on the medium and fast trajectories, dominated by lag-residual jitter that the fixed per-shape $N^{*}$ compensation cannot eliminate in this regime.

\paragraph{Graceful versus catastrophic degradation beyond the design envelope.}
The $-25$\,dB cells lie $5$\,dB below both the training envelope and the design range of every published baseline, and below the Schottky detector chain's specified operating range; we report them as a degradation-mode characterisation, not as an operating-range comparison. Read that way, the stress test separates one graceful degrader from six collapses. The deep-learning estimator's geometric mean rises from the corresponding $-20$\,dB cells of Table~\ref{tab:pure_runs} by about a factor of eight under the $5$\,dB noise-power increase: degraded, but still tracking the trajectory for the majority of frames on every shape. Every alternative effectively loses the beam, with miss fractions approaching or exceeding two thirds: the classical estimator's two-frame motion-compensated pool carries too little evidence for the matched-filter argmax to find the sideband at all, and the three stateless baselines sit at the uniform-error level of a readout dominated by noise maxima (Figure~\ref{fig:ood_snr}). The deep-learning estimator's margin here is the same mechanism as its $-20$\,dB lead (a full-spectrum likelihood integrated across frames by the $(q, v)$ posterior), pushed past the point where every per-frame-decision architecture breaks.

\paragraph{Operating-regime concentration of the advantage.}
Taken together, these out-of-distribution tests sharpen the operating-regime picture of Table~\ref{tab:pure_runs} and Section~\ref{subsec:narrowband} into a single SNR axis. The deep-learning estimator is the strongest estimator at and below $-10$\,dB (every in-distribution cell there and every $-25$\,dB stress cell), which is exactly the regime a diagnostics chain cannot choose to avoid. From $-5$\,dB upward the CNN+KF baseline's calibrated scalar head is consistently the most accurate (by $10$--$50\%$ over the deep-learning estimator), but in that regime every temporal estimator sits within a factor of a few of the quantisation reference and the absolute errors are operationally comfortable, so the high-SNR ordering carries little deployment weight. The classical estimator's PSD-domain pooling and matched-filter readout remain competitive on slow trajectories and stationary spectra across the mid and high SNRs, and it remains the strongest CPU-only, ML-free option; its weaknesses concentrate at the deep-SNR edge and under drifting clutter (Section~\ref{subsec:narrowband}). The deep-learning estimator is therefore the dependable default wherever the SNR budget is tight or unpredictable, while above $-5$\,dB the choice among the temporal estimators is better made on latency, hardware, and robustness grounds than on accuracy.

\subsection{Component ablations and system characterisation}
\label{subsec:ablation}
\subsubsection{Classical-estimator ablation}

\paragraph{Motion-compensated temporal pooling.}
The classical estimator carries its temporal context in a single block, the motion-compensated coherent EMA of Section~\ref{subsec:emasc}; the postprocess smoother, the matched-filter bank, and the readout are stateless within a frame. We ablate the two design decisions of this block, the temporal pooling itself and its velocity-shifted accumulation, on the $15$ broadband-noise cells of Section~\ref{subsec:pure_runs} (the same frame draw as Table~\ref{tab:pure_runs}), with the estimator at its tune-unit window setting of Section~\ref{subsubsec:input_length}. The \emph{full} configuration is the deployed coherent EMA; \emph{EMA disabled} sets the decay to zero, so the matched-filter bank reads the single postprocessed frame; \emph{EMA without motion compensation} keeps the same decay but drops the per-frame velocity shift, accumulating successive frames at their raw bin positions.

\begin{table}[htbp]
\centering
\caption{Classical-estimator ablation of the motion-compensated coherent EMA. Geometric-mean MAE in $10^{-3}$ tune units over the $15$ broadband-noise cells of Table~\ref{tab:pure_runs} (same frame draw), with the estimator in its deployed configuration. \emph{Full}: the deployed coherent EMA. \emph{EMA disabled}: decay $=0$, so the matched-filter bank reads the single postprocessed frame. \emph{EMA without motion compensation}: the same decay with no velocity-shifted accumulation. Bold marks the smallest value. \label{tab:emasc_ablation}}
\smallskip
\footnotesize
\begin{tabular}{l c}
\toprule
configuration & geometric-mean MAE \\
\midrule
full (deployed)                 & \textbf{0.418} \\
EMA disabled                    & 1.283          \\
EMA without motion compensation & 0.961          \\
\bottomrule
\end{tabular}
\end{table}

Table~\ref{tab:emasc_ablation} reports the ablation, and both decisions are load-bearing. Disabling the temporal pooling raises the broadband-noise geometric mean by a factor of $3.1$, with the largest penalty at the deepest SNR where single-frame evidence is weakest: on these single-acquisition-noise frames the matched-filter bank cannot localise the sideband reliably from one frame alone. Removing the velocity shift while keeping the pooling costs a factor of $2.3$: a motion-uncompensated EMA smears the moving sideband across the bins it visits as the tune evolves, surrendering most of the pooling gain, though even smeared pooling retains an edge over no pooling at this noise level. The two mechanisms therefore compose: cross-frame pooling supplies the noise averaging, and the velocity-shifted realignment is what converts that averaging into a sharp, correctly-located sideband. The full configuration reproduces the classical column of Table~\ref{tab:pure_runs} on the same frame draw.

\subsubsection{Deep-learning-estimator ablation}

\begin{table}[htbp]
\centering
\caption{Deep-learning-estimator ablation: mean absolute tune error in $10^{-3}$ tune units across the $15$ broadband-noise cells, for readout and tracker variants on the two estimators' CNN backbones. The slow, medium, and fast columns are geometric means over the five SNR levels of that trajectory shape. The deep-learning backbone with the deployed $(q, v)$ Bayes tracker is the deployed estimator; its Kalman variant feeds the peak-local soft-argmax mean and variance of the same map to a fair $(q, v)$ constant-velocity Kalman filter with the deployed motion constants; the $(q)$-only Bayes variant is the deployed tracker with the velocity envelope set to zero. The CNN+KF backbone with a Kalman filter is the CNN+KF baseline of Table~\ref{tab:pure_runs}. $T_{5 \cdot 10^{-3}}$ is the mean per-cell catastrophic-miss rate, in percent. Bold marks the smallest geometric mean. \label{tab:dl_ablation}}
\smallskip
\footnotesize
\setlength{\tabcolsep}{5pt}
\begin{tabular}{l l c c c c c}
\toprule
CNN backbone & readout / tracker & slow & medium & fast & geomean & $T_{5 \cdot 10^{-3}}$ (\%) \\
\midrule
\multirow{4}{*}{deep-learning} & no tracker (soft-argmax) & 0.690 & 0.684 & 0.682 & 0.685 & 9.05 \\
 & Kalman (peak-local mean$+$var) & 0.421 & 0.390 & 0.447 & 0.418 & 5.30 \\
 & $(q, v)$ Bayes (deployed) & 0.164 & 0.164 & 0.245 & \textbf{0.187} & 0.121 \\
 & $(q)$-only Bayes & 2.594 & 5.488 & 16.207 & 6.133 & 39.82 \\
\midrule
\multirow{3}{*}{CNN+KF} & no tracker (CNN-only) & 0.412 & 0.393 & 0.379 & 0.394 & 4.00 \\
 & Kalman ($=$ CNN+KF) & 0.189 & 0.222 & 0.338 & 0.242 & 1.77 \\
 & $(q, v)$ Bayes & 0.117 & 0.184 & 0.615 & 0.237 & 0.343 \\
\bottomrule
\end{tabular}
\end{table}

\paragraph{Tracker and CNN-backbone decomposition.}
Table~\ref{tab:dl_ablation} decomposes the deep-learning estimator into its CNN backbone and its temporal tracker, holding the deployed checkpoint and preprocessing fixed. Temporal tracking is load-bearing: on the deep-learning backbone the no-tracker soft-argmax reads about $3.7\times$ the deployed tracker, and its per-frame catastrophic-miss rate falls to a rare event once the tracker accumulates evidence across frames. Part of this per-frame gap is structural rather than a map-quality deficit: the deployed network is trained with a windowed likelihood objective that leaves the probability mass far from the sideband unconstrained, so a whole-map soft-argmax readout is penalised by design, whereas the tracker consumes the same raw map as a per-frame likelihood and is unaffected. What the tracker consumes matters as much as that it tracks: the Kalman variant on the same backbone (which reduces each map to a peak-local mean and variance before filtering) reads $2.2\times$ the deployed tracker with a far heavier catastrophic-miss tail, because a scalar reduction commits to the dominant peak before filtering, and at deep SNR the committed peak is wrong on a few percent of frames, errors no Gaussian measurement model can subsequently reject. The discrete $(q, v)$ posterior consumes the full multi-modal map instead, so a wrong dominant peak on one frame remains a recoverable hypothesis rather than a corrupted measurement. On the CNN+KF backbone, whose scalar regression head is its native output, the same comparison is nearly tied in the mean with a five-fold tail reduction for the discrete posterior, consistent with the same mechanism operating on a much better-behaved scalar. Cross-pairing the components is, however, not a free exchange: the swapped pair's gains concentrate on slow trajectories and reverse on fast ones, and the deployed pairing beats the swapped one on both the mean and the tail, so the deep-learning estimator's accuracy is a property of the CNN and the tracker trained and configured together rather than of either alone. The velocity coordinate itself is a precondition rather than a refinement: with the velocity envelope zeroed, the motion prior pulls the posterior toward the previous tune position instead of toward where the sideband has moved, and tracking collapses to $33\times$ the deployed tracker, with catastrophic misses on a large fraction of frames and the damage steepest on fast trajectories. This component decomposition feeds the deployment-context trade-offs of the decision matrix (Section~\ref{subsec:decision_matrix}).

\paragraph{Cross-validation reliability.}
To confirm that the reported deep-learning-estimator accuracy is not an artefact of a single training split, we evaluate all five cross-validation fold checkpoints of Section~\ref{subsec:setup} on the $9$ broadband-noise cells (the three trajectory shapes at $-10$, $-15$, and $-20$\,dB). The $9$-cell geometric-mean MAE across the five folds is $0.284 \pm 0.006 \times 10^{-3}$ tune units, a spread of a few percent about the mean with the deployed checkpoint among the tightly clustered folds. This confirms the reported accuracy is robust to the training split rather than a favourable-split selection. This is a generalisation check rather than a component ablation: it varies the training split, not the estimator architecture.

\subsubsection{Latency and hyperparameter portability}

\paragraph{Per-frame latency.}
Table~\ref{tab:latency} reports the per-frame inference latency of both proposed estimators at the deployed spectrum length $L = 1024$, the SAPT operating configuration, with the classical estimator on a single CPU core and the deep-learning estimator on a consumer-class GPU at batch size~$1$. Each estimator is timed per frame over several thousand frames, after a warm-up that absorbs the one-time initialisation costs (just-in-time compilation for the classical estimator; Compute Unified Device Architecture (CUDA) graph capture and kernel compilation for the deep-learning estimator), and we report the median and $99$-th percentile. The reported latencies are those of the downstream estimator alone; the shared upstream preprocessing (Welch periodogram formation, revolution-frequency estimation, folded-grid mapping, and, for real-beam data, channel eigendecoupling) is common to all estimators and is excluded, so the comparison isolates the estimators rather than the full acquisition-to-readout chain. At this deployed length both estimators satisfy the SAPT per-frame latency budget.

\begin{table}[htbp]
\centering
\caption{Per-frame inference latency of the two proposed estimators at the deployed length $L = 1024$. The classical estimator runs on a single CPU core; the deep-learning estimator runs on a single consumer-class GPU at batch size~$1$, with its entire per-frame chain (feature build, likelihood-map CNN, and $(q, v)$ tracker update) replayed as one captured CUDA graph. \label{tab:latency}}
\smallskip
\footnotesize
\begin{tabular}{l l c c}
\toprule
estimator & deployment & median ($\mu\mathrm{s}$) & p99 ($\mu\mathrm{s}$) \\
\midrule
classical & CPU & $115$ & $328$ \\
deep-learning & GPU (CUDA graph) & $397$ & $823$ \\
\bottomrule
\end{tabular}
\end{table}

The two estimators sit in different latency bands, both with sub-millisecond median and $99$-th-percentile per-frame latency, within the SAPT inter-frame budget. The classical estimator runs on a single CPU core; the deep-learning estimator runs on a commodity GPU, where capturing its whole per-frame chain as a single CUDA graph is what brings it into budget: executed eagerly, one kernel launch at a time, the same chain is about an order of magnitude slower (a median near $4$\,ms, above the inter-frame budget), so the graph capture is a deployment requirement rather than a mere optimisation. The latency split is one of the cleanest deployment-context discriminators between the two estimators: a control rack without GPU support cannot run the deep-learning estimator within the latency budget, while a control rack with a commodity GPU can run either; we return to this point in Section~\ref{subsec:decision_matrix}.

\paragraph{Tracker hyperparameter portability.}
The deep-learning estimator's tracker hyperparameters fall into two groups with different portability profiles. The velocity-grid parameters $(V_{\max}, \sigma_v, \rho)$ are tied to the per-frame tune-step distribution induced by the facility's frame rate and ramp dynamics, and would be re-derived from operational specifications when the estimator is transferred to a machine with substantially different timing. The remaining tracker parameter, the floor-residual $q$-axis process noise $\sigma_q = 1/\sqrt{12}$\,bins, is an algorithmic stability margin on the discrete grid: it is a discrete-state-space constant tied to the integer-shift quantisation rather than to any physical scale (it scales with the velocity-grid stride when that latency lever is engaged, one lattice quantum per step), and can be re-used unchanged across facilities and grid resolutions. Because none of the deep-learning-estimator tracker hyperparameters are learned weights (they are explicit configuration entries or fixed algorithmic constants such as $\sigma_q$), recalibration for a new facility amounts to a configuration-level change and does not require retraining the network. The classical-estimator hyperparameters follow the same pattern: the grid-dependent scales are explicit configuration entries expressed in physical tune units (Section~\ref{subsec:emasc}), hence shared across grid sizes, while the remainder are dimensionless constants (polynomial order, decay factor, velocity-history length); there are no learned weights to recalibrate.

\subsubsection{Input-length scaling}
\label{subsubsec:input_length}

Both proposed estimators are defined once for all spectrum lengths: every window, velocity, and process-noise parameter is specified in tune units and converted to bins on the operating grid (Sections~\ref{subsec:emasc} and~\ref{subsec:nlbf}), and the deep-learning network is fully convolutional. This subsection characterises what a longer input spectrum actually buys. Unless stated otherwise, entries are geometric-mean MAE in $10^{-3}$ tune units over the operating-range cells of the dynamic-tune grid (three trajectory shapes $\times$ SNR $\in \{-15, -10, -5, 0\}$\,dB; the $-20$\,dB edge cells are reported alongside in the released per-cell records). For these runs the classical estimator uses its deployed tune-unit windows $w_q = W_{\mathrm{SG}} = 10^{-2}$ (Section~\ref{subsec:emasc}), converted to bins on each operating grid, so the same configuration is evaluated at every $L$ and in the fixed-$L$ benchmarks of the preceding sections.

\begin{table}[htbp]
\centering
\caption{Input-length scaling at the coarsest acquisition resolution ($\Delta f \approx 5\,\mathrm{kHz}$). Geometric-mean MAE in $10^{-3}$ tune units over the operating-range cells (SNR $\ge -15$\,dB) of the dynamic-tune grid, per input length $L$. \emph{DL, no tracker}: the deep-learning map read per frame by the whole-map soft-argmax; \emph{DL, tracker}: the deployed deep-learning estimator; \emph{classical}: the tune-unit-parameterised classical estimator. Bold marks the best value per column. \label{tab:input_length}}
\smallskip
\footnotesize
\begin{tabular}{l c c c c c}
\toprule
estimator & $L{=}512$ & $L{=}1024$ & $L{=}2048$ & $L{=}4096$ & $L{=}8192$ \\
\midrule
DL, no tracker & 0.371 & 0.279 & 0.273 & 0.300 & 0.374 \\
DL, tracker (deployed) & \textbf{0.169} & \textbf{0.141} & \textbf{0.127} & \textbf{0.128} & \textbf{0.146} \\
classical & 0.196 & 0.217 & 0.225 & 0.208 & 0.199 \\
\bottomrule
\end{tabular}
\end{table}

Table~\ref{tab:input_length} sweeps $L$ at the coarsest acquisition resolution, $\Delta f \approx 5$\,kHz, where the information content of the input is fixed by the acquisition and a longer spectrum only re-grids it. All three arms are, correctly, nearly flat: no estimator can convert grid points into accuracy that the acquisition does not carry. The deep-learning tracker's mild optimum at $L = 2048$ follows the per-frame map's own quality ordering (the no-tracker row), and the tracker improves on its map by the same factor of roughly $2.2$ at every length; the classical estimator stays within $\pm 8\%$ of its mean across the full $16\times$ length range. The ordering between the arms is also length-independent: the deployed tracker leads the classical estimator everywhere, and the classical estimator leads the per-frame map readout everywhere. At the $-20$\,dB edge the gap between the deep-learning tracker and the classical estimator widens to a factor of $8$ to $10$, the same low-SNR separation seen throughout Section~\ref{subsec:robustness}.

\subsubsection{Acquisition-resolution grid}
\label{subsubsec:acq_resolution}

\begin{table}[htbp]
\centering
\caption{Acquisition-resolution grid: geometric-mean MAE in $10^{-3}$ tune units over the operating-range cells (SNR $\ge -15$\,dB), per acquisition resolution $\Delta f$ and input length $L$. Upper panel: the deployed deep-learning estimator; lower panel: the tune-unit-parameterised classical estimator. Bold marks the best $L$ per column. \label{tab:acq_resolution}}
\smallskip
\footnotesize
\setlength{\tabcolsep}{5.5pt}
\begin{tabular}{l c c c c c}
\toprule
 & $\Delta f{\approx}256$\,Hz & $512$\,Hz & $1024$\,Hz & $2048$\,Hz & $4096$\,Hz \\
\midrule
\multicolumn{6}{l}{\emph{deep-learning, tracker (deployed)}} \\
$L{=}512$  & 0.083 & 0.092 & 0.103 & 0.119 & 0.151 \\
$L{=}1024$ & 0.044 & 0.049 & 0.059 & 0.079 & 0.121 \\
$L{=}2048$ & 0.021 & 0.026 & 0.038 & \textbf{0.065} & \textbf{0.110} \\
$L{=}4096$ & 0.012 & \textbf{0.019} & \textbf{0.035} & 0.066 & \textbf{0.110} \\
$L{=}8192$ & \textbf{0.010} & 0.021 & 0.044 & 0.078 & 0.127 \\
\midrule
\multicolumn{6}{l}{\emph{classical}} \\
$L{=}512$  & 0.100 & 0.110 & 0.125 & 0.142 & 0.177 \\
$L{=}1024$ & 0.092 & 0.103 & \textbf{0.117} & \textbf{0.139} & 0.192 \\
$L{=}2048$ & 0.092 & 0.103 & 0.121 & 0.153 & 0.207 \\
$L{=}4096$ & \textbf{0.090} & \textbf{0.103} & 0.128 & 0.164 & 0.198 \\
$L{=}8192$ & \textbf{0.090} & 0.112 & 0.138 & 0.158 & \textbf{0.186} \\
\bottomrule
\end{tabular}
\end{table}

Table~\ref{tab:acq_resolution} completes the picture by sweeping the acquisition resolution $\Delta f$ jointly with $L$, on the same frame streams and trajectories for every cell. Three structural observations follow. First, at fine acquisition resolution the deep-learning estimator converts length directly into accuracy: at $\Delta f \approx 256$\,Hz its error falls monotonically by a factor of eight from $L = 512$ to $L = 8192$, tracking the $\propto 1/L$ scaling of its per-frame map with a diminishing final step ($-12\%$ from $L = 4096$ to $L = 8192$) as the acquisition's information floor is approached. Second, the optimal length tracks the acquisition bandwidth: the best $L$ per column moves from $8192$ at $\Delta f \approx 256$\,Hz to $2048$--$4096$ at $\Delta f \ge 2048$\,Hz, an information-matching rule of thumb $L_{\mathrm{opt}} \approx f_{\mathrm{rev}} / (2\,\Delta f)$ under which a deployment should choose the grid that matches its acquisition rather than the largest one it can afford. Third, the classical estimator benefits from finer acquisition ($\times 2$ from $\Delta f \approx 4096$ to $256$\,Hz at fixed $L$, as box-average decimation noise recedes) but is flat in $L$ at every $\Delta f$ (within $10\%$ from $L = 512$ to $L = 8192$ at $\Delta f \approx 256$\,Hz): its argmax-plus-centroid readout localises the peak to a fixed fraction of the sideband width regardless of how finely that sideband is gridded. The two panels thereby separate the two resolution resources: acquisition bandwidth helps both estimators, while grid length is convertible into accuracy only by the estimator whose posterior sharpens with the grid, and the deep-learning tracker's lead over the classical arm grows from $1.2\times$ at $L = 512$ to $8.7\times$ at $L = 8192$ under the finest acquisition. This is the quantitative basis for the variable-resolution generalist claim of Section~\ref{sec:conclusions}.

\subsection{Preliminary validation on operational beam data}
\label{subsec:real_beam}
As a preliminary validation under operational conditions, we apply both proposed estimators, together with the CNN+KF~\cite{sun2026real} and latency-compensated T-PD~\cite{sun2025high} baselines as further independent estimators, to real-beam data recorded at the SAPT facility on two detectors: an existing beam-position monitor (BPM) and a purpose-built Schottky-monitor prototype. The BPM recording is the same operational recording analysed in the companion CNN+KF study~\cite{sun2026real}, here re-analysed under the four-estimator comparison of the present work. It consists of $20\,480\,000$ samples per BPM electrode at a sampling rate of $f_s = 29.75$\,MHz, captured in a single continuous acquisition with no externally controlled SNR floor. The Schottky-monitor prototype contributes ten separate coasting-beam acquisitions at $f_s = 50$\,MHz spanning the $70$--$230$\,MeV energy range, likewise with no externally controlled SNR floor. Unlike the synthetic benchmarks of Sections~\ref{subsec:pure_runs} to~\ref{subsec:ablation}, the operating tune of these acquisitions is not separately known; we therefore treat the median across the three deployable estimators (classical, CNN+KF, deep-learning) as a model-agnostic consensus reference and report stability and cross-estimator agreement, deferring validation under actual operating conditions to the SAPT commissioning campaign.

\begin{figure}
    \centering
    \includegraphics[width=\linewidth]{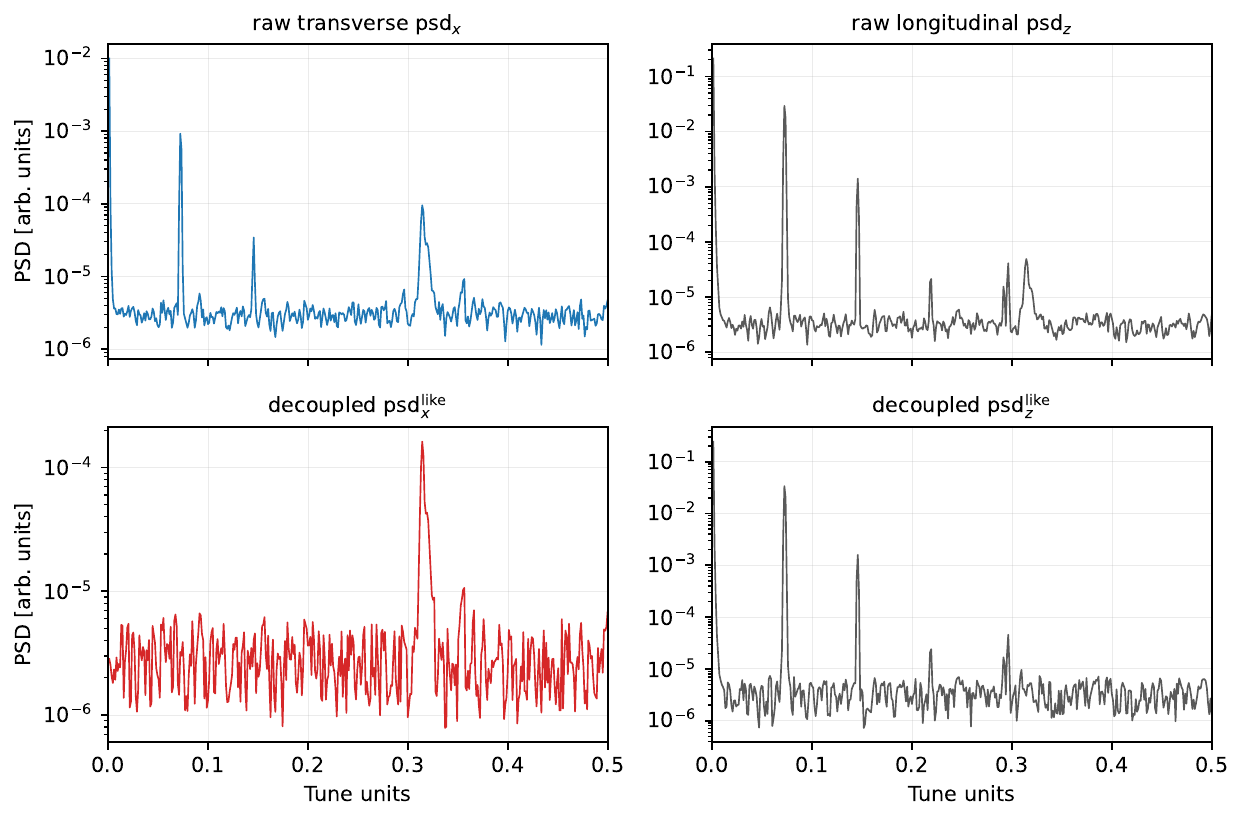}
    \caption{Effect of the eigendecoupling on a representative real-beam frame. The horizontal axis is the betatron tune, obtained by mirroring the upper half-band ($f / f_{\mathrm{rev}} \in [0.5, 1)$) onto $[0, 0.5)$ via $\mathrm{tune} = 1 - f / f_{\mathrm{rev}}$ at the native FFT resolution. Top row: power spectral densities of the raw transverse-difference channel $\mathrm{psd}_x$ (left) and the raw longitudinal-sum channel $\mathrm{psd}_z$ (right) before decoupling. Bottom row: the decoupled transverse projection $\mathrm{psd}_x^{\mathrm{like}}$ (left) and longitudinal channel $\mathrm{psd}_z^{\mathrm{like}}$ (right) after decoupling. The narrow-band interference lines that overlap the transverse band in the raw signal are largely removed from $\mathrm{psd}_x^{\mathrm{like}}$ and concentrated in the longitudinal channel, leaving the betatron sideband near tune $\approx 0.32$ as the dominant transverse feature for the downstream tune extraction.}
    \label{fig:decouple_psd}
\end{figure}

\paragraph{Decoupling and frame construction.}
The four BPM electrodes $a$, $b$, $c$, $d$ are arranged around the beam pipe so that the transverse position of the beam centroid couples linearly into the electrode signals through small geometric coefficients, while longitudinal beam-current modulation couples into all four electrodes through a larger common-mode coefficient. In raw form, both transverse-direction differential sums and the longitudinal sum can therefore contain a substantial longitudinal contribution, which has a different Schottky-spectrum fingerprint but overlaps the same frequency band. We construct three orthogonal channels by linear combinations,
\begin{align}
\label{eq:decouple-x}
    X &= a + d - b - c,\\
\label{eq:decouple-y}
    Y &= a + b - c - d,\\
\label{eq:decouple-z}
    Z &= a + b + c + d,
\end{align}
which correspond respectively to horizontal-difference, vertical-difference, and longitudinal-sum modes of the four-electrode array. Even after these linear combinations, residual lateral cross-talk and longitudinal contamination can still contribute to the transverse channels because of small mechanical and electrical asymmetries between the electrodes. To isolate the dominant horizontal transverse component, we apply a per-frequency eigendecoupling step on the $3\times3$ cross-spectral density matrix between $X$, $Y$, and $Z$. The three eigenvectors at each frequency define a basis in which the three modes are uncorrelated, and the residual cross-talk that remains after the linear combinations is concentrated into one of the three eigen-modes rather than spread across all three. We assign each eigenvalue/eigenvector pair to the horizontal-, vertical-, or longitudinal-like axis by maximising the projection of the eigenvector onto each linear combination, and use the horizontal-like spectrum $\mathrm{psd}_x^{\mathrm{like}}$ as the input to the downstream tune extraction. In the representative frames shown below, narrow-band interference lines that originally appeared in the transverse spectrum are largely assigned to the longitudinal channel $\mathrm{psd}_z$, where they no longer dominate the betatron-tune estimate. Figure~\ref{fig:decouple_psd} compares the transverse and longitudinal spectra before and after the decoupling step on representative frames; the suppression of narrow-band lines in the transverse channel and their concentration in the longitudinal channel is the operational evidence that the decoupling acts as intended. The standard low-tune mask of the feature builder is also applied here, although its operational role is reduced because the decoupling already removes most of the low-frequency leakage upstream.

The existing-BPM recording, taken on a coasting beam at a kinetic energy of approximately $142$\,MeV ($f_{\mathrm{rev}} \approx 6$\,MHz), is partitioned into $625$ non-overlapping frames of $32\,768$ points per frame, each spanning an acquisition window of approximately $1.10$\,ms; within each such frame a Welch periodogram is computed using $4096$-point Hann-windowed segments. The ten Schottky-monitor-prototype acquisitions are partitioned and Welch-averaged following the same procedure. In both cases the revolution frequency is estimated as the strongest peak of $\mathrm{psd}_z$ above $1$\,MHz, and the resulting normalised frequency axis is mapped onto the standard $1024$-bin tune grid using the same folded-grid mapping after normalisation by the revolution-frequency estimate.

\paragraph{Stability and cross-estimator agreement.}
Table~\ref{tab:real_beam} reports the per-estimator stability across the full real-beam set, pooling the existing-BPM recording ($625$ frames at $\hat q \approx 0.316$) with the ten Schottky-monitor-prototype acquisitions spanning $70$--$230$\,MeV (operating points $\hat q = 0.20$--$0.32$). No estimator produced divergent or non-finite output on any acquisition, and the deployable-estimator per-acquisition medians lock onto the same operating point at every energy, agreeing to within $2.5 \times 10^{-3}$. Because the acquisitions sit at different operating points, every dispersion metric is computed within each acquisition and then pooled, so it measures frame-to-frame stability independently of the absolute tune. All of these acquisitions are coasting beam; the acceleration-ramp windows are not covered, owing to limitations of the present data-acquisition system.

\begin{table}[htbp]
\centering
\caption{Real-beam stability across the two SAPT detectors, pooled over the existing-BPM recording ($625$ frames at $\hat q \approx 0.316$) and ten Schottky-monitor-prototype acquisitions spanning $70$--$230$\,MeV, whose deployable-estimator consensus operating point ranges over $\hat q = 0.20$--$0.32$ and agrees to within $2.5\times10^{-3}$ at every energy. Values are in $10^{-3}$ tune units. Because the acquisitions sit at different operating points, every dispersion is computed within each acquisition and then pooled, so it measures frame-to-frame stability independently of the absolute tune. ``std'' is the pooled within-acquisition standard deviation of the estimator's own output; ``jitter'' is the RMS of frame-to-frame differences; ``mean $\hat\sigma$'' is the mean reported posterior standard deviation; ``mean $|\Delta_{\mathrm{med}}|$'' is the mean absolute deviation from the per-frame median across the three deployable estimators (classical, CNN+KF, deep-learning). The latency-compensated T-PD baseline of~\cite{sun2025high} is shown for comparison and is not folded into the consensus. The classical and T-PD estimators do not report a posterior $\hat\sigma$. \label{tab:real_beam}}
\smallskip
\begin{tabular}{l c c c c}
\toprule
estimator & \multicolumn{1}{c}{std} & \multicolumn{1}{c}{jitter} & \multicolumn{1}{c}{mean $\hat\sigma$} & \multicolumn{1}{c}{mean $|\Delta_{\mathrm{med}}|$} \\
\midrule
deep-learning (proposed) & 0.466 & 0.362 & 0.451 & 0.106 \\
classical (proposed) & 0.493 & 0.304 & \multicolumn{1}{c}{n.a.} & 0.215 \\
CNN+KF~\cite{sun2026real}        & 6.132 & 5.523 & 0.264 & 1.349 \\
T-PD~\cite{sun2025high}          & 0.372 & 0.168 & \multicolumn{1}{c}{n.a.} & 0.570 \\
\bottomrule
\end{tabular}
\end{table}

\begin{figure}[htbp]
\centering
\includegraphics[width=\linewidth]{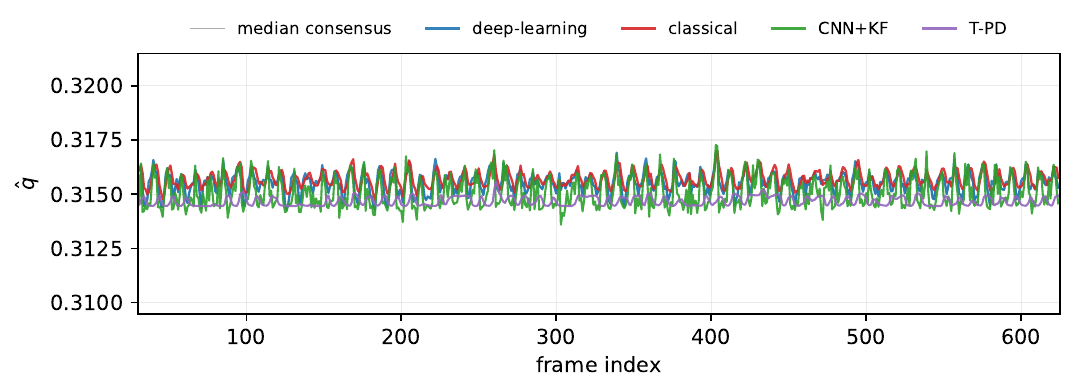}
\caption{Per-frame tune estimate $\hat q_t$ of the four estimators across the frames of the SAPT BPM recording, the near-stationary component of the real-beam set of Table~\ref{tab:real_beam}. The three deployable estimators (deep-learning, classical, CNN+KF) track the same operating point at $\hat q \approx 0.3155$, the grey curve being their per-frame median consensus; the latency-compensated T-PD baseline sits about $0.9 \times 10^{-3}$ below the consensus with the lowest frame-to-frame jitter. \label{fig:real_beam_trace}}
\end{figure}

The deep-learning estimator achieves the tightest cross-estimator agreement over the full set, ahead of the classical estimator and the two baselines. These statistics measure deviation from the cross-method consensus rather than from a known tune, and should be read as a check that the deployable estimators converge on the same operating point on real beam, not as an accuracy ranking. The BPM recording sits at a near-stationary operating point with low contaminant pressure, where all four estimators track within noise. The Schottky-prototype acquisitions are less benign: the $200$\,MeV records in particular expose a per-frame failure mode of the CNN+KF baseline, whose per-frame regression head, unlike the temporally-integrated or full-spectrum estimators, carries no cross-frame outlier gate. Although its Kalman-tracked median locks onto the correct tune at every energy, a few percent of its per-frame readouts jump to spurious folded peaks, inflating its pooled dispersion and cross-estimator deviation by an order of magnitude while its median-based agreement stays tight. The proposed estimators and the T-PD baseline reject these frames through full-spectrum or cross-frame evidence, the spectral-ambiguity mechanism that separates them from per-frame architectures in the synthetic studies of Section~\ref{subsec:pure_runs}; the velocity-aware tracking, by contrast, remains untested on this coasting-beam data.

\paragraph{Outlook.}
These acquisitions are coasting beam at near-stationary operating points and do not exercise the full operating envelope of the SAPT acceleration-ramp and extraction windows. The velocity-aware tracking that distinguishes the proposed estimators in the dynamic-tune regime is not exercised by this coasting-beam data, and validation of the proposed estimators' accuracy and real-time measurement capability under actual operating conditions is deferred to the SAPT commissioning campaign. The preliminary validation reported here establishes that both proposed estimators run end-to-end without divergence and track a consistent operating point across both detectors, as quantified by the stability and cross-estimator agreement results above.

\subsection{Deployment-oriented guidance}
\label{subsec:decision_matrix}
The main dynamic-tune benchmark (Section~\ref{subsec:pure_runs}), the beam-loss stress test (Section~\ref{subsec:beam_loss}), the narrow-band robustness study (Section~\ref{subsec:narrowband}), the component ablations and resolution studies (Section~\ref{subsec:ablation}), and the SAPT real-beam validation (Section~\ref{subsec:real_beam}) together identify a consistent pattern. The deep-learning estimator is the stronger of the two proposed estimators wherever the conditions are hard: it wins every broadband-noise cell in the low-SNR regime and the aggregate geometric mean, leads the drifting-clutter regime, and is the only estimator that degrades gracefully below the design envelope; in the higher-SNR regime the published CNN+KF baseline is consistently the most accurate, where every temporal estimator's absolute error is operationally comfortable. The classical estimator's complementary advantage is deployment economy rather than accuracy: it runs on a single CPU core with no GPU and no training data, and stays competitive at moderate SNR on benign, slowly-varying operating points. The choice is therefore governed by deployment context rather than by a single accuracy ranking. This subsection collects that recommendation into Table~\ref{tab:decision_matrix} and discusses the deployment-context dimensions (training-data availability and posterior uncertainty) that the per-cell accuracy sections of Sections~\ref{subsec:pure_runs} to~\ref{subsec:ablation} do not separately cover.

\begin{table}[htbp]
\centering
\caption{Deployment-oriented decision matrix for the two proposed estimators. Each row is an operating-regime or deployment-context dimension along which the choice between the classical estimator and the deep-learning estimator is informed by the empirical results of Sections~\ref{subsec:pure_runs} to~\ref{subsec:real_beam}. \label{tab:decision_matrix}}
\smallskip
\footnotesize
\setlength{\tabcolsep}{4pt}
\begin{tabular}{p{2.2cm} p{3.6cm} p{3.6cm} p{1.7cm}}
\toprule
dimension & classical preferred & deep-learning preferred & evidence \\
\midrule
accuracy & strong, well-resolved sideband; benign spectra & weak or buried sideband; degraded spectra & Tab.~\ref{tab:pure_runs}, \ref{tab:highsnr} \\
\midrule
robustness & faster re-lock after beam loss & drifting clutter; below-envelope SNR & Tab.~\ref{tab:beam_loss}, \ref{tab:narrowband} \\
\midrule
hardware & single CPU core, no GPU & commodity GPU & Tab.~\ref{tab:latency} \\
\midrule
resolution (input length $L$) & any $L$; accuracy flat in $L$ & any $L \in [512, 8192]$; finer $L$ improves accuracy & Tab.~\ref{tab:input_length}, \ref{tab:acq_resolution} \\
\midrule
training data & none required & training set + compute & Sec.~\ref{subsec:nlbf} \\
\midrule
latency & sub-$0.4$\,ms p99 (CPU) & sub-$0.9$\,ms p99 (GPU) & Tab.~\ref{tab:latency} \\
\bottomrule
\end{tabular}
\end{table}

\paragraph{Training-data availability.}
The classical estimator has no learnable parameters and is shipped with the same hyperparameter values across facilities; the recalibration cost when transferring to a new accelerator is a configuration-level change. The deep-learning estimator requires a training set with the operating-condition coverage described in Section~\ref{subsec:preprocessing} and the cross-fold protocol of Section~\ref{subsec:setup}. This is the deployment-context dimension along which the two estimators differ most cleanly; it is also the dimension along which the comparison paper should be read as a service to the practitioner rather than as an accuracy-ranking exercise.

\paragraph{Posterior $\hat\sigma$ availability.}
The deep-learning estimator emits a per-frame posterior $\hat\sigma_t$ derived from the marginal posterior over the discrete state grid, providing an uncertainty channel for downstream gating; a control loop using it should carry a small external margin, since the reported width can run below the realised dispersion under un-modelled operating-point motion. The classical estimator emits a per-frame tune estimate but does not emit a posterior standard deviation; downstream gating that requires an uncertainty channel must be supplied externally (e.g., by frame-level signal-quality features upstream of the estimator) when the classical estimator is used.

\paragraph{Combined recommendation.}
The decision matrix admits two clean default deployments and one mixed deployment. A clinical accelerator without GPU support and with confirmed operating conditions in the moderate-SNR regime should default to the classical estimator, whose accuracy there is competitive on the trajectories such a machine actually runs and whose deployment cost is a single CPU core; the caveats are the deep-SNR edge and drifting clutter, where it degrades more than the deep-learning estimator. A facility with GPU support should default to the deep-learning estimator: it is the most accurate estimator wherever accuracy is scarce (the low-SNR, degraded-spectrum regime), the most robust under drifting clutter and below-envelope SNR, meets the latency budget, and, without retraining, converts a longer input spectrum into finer tune accuracy over the evaluated length range, whereas the classical estimator's accuracy stays flat with input length (the resolution sweeps cover the two proposed estimators only). A facility with GPU support but with a desire for an interpretable, deployment-fixed pipeline as a backup may run both estimators in parallel, with the per-frame agreement between the two as an additional integrity signal; the cross-estimator deviation on the BPM recording (Section~\ref{subsec:real_beam}) demonstrates that this kind of cross-check produces useful agreement statistics on operational data without a known ground-truth tune.

\section{Conclusions}
\label{sec:conclusions}
This work addressed the per-frame estimator that consumes the normalised folded-tune spectrum of~\cite{sun2025high,sun2026real}, with both proposed estimators parameterised in tune units so that the grid length is a free configuration rather than a constraint and they run at an arbitrary $L$ within a wide range: once the upstream chain is fixed, the dominant tune-readback error originates in this layer. We proposed two complementary estimators that share the spectral front-end but carry temporal context through structurally different representations. The classical estimator (Section~\ref{subsec:emasc}) pools spectral evidence across frames through a motion-compensated coherent moving average and reads the tune with a matched-filter bank and a gated sub-bin centroid; the deep-learning estimator (Section~\ref{subsec:nlbf}) maps each spectrum to a tune-grid likelihood with a convolutional network and propagates a discrete $(q, v)$ Bayes posterior under a Gaussian motion model.

On the synthetic dynamic-tune benchmark of Section~\ref{subsec:pure_runs} the deep-learning estimator achieves the best aggregate accuracy, ahead of both the published CNN+KF and latency-compensated T-PD baselines of~\cite{sun2026real,sun2025high}, with the classical estimator between them. Its advantage concentrates where accuracy is scarce: it leads across the low-SNR regime, carries the lowest catastrophic-miss rate by a wide margin, is the most robust under drifting narrow-band clutter, and is the only estimator that degrades gracefully below the design envelope. It is also the one that converts finer acquisition resolution into accuracy as the spectrum lengthens, while the classical estimator, equally length-portable after the same tune-unit parameterisation, is flat in length. The classical estimator's complementary value is deployment economy: a single CPU core, no training-data dependency, and competitive accuracy at moderate SNR on benign operating points. Both meet the real-time per-frame budget, on a single CPU core and a commodity GPU respectively, and the operating-regime decision matrix of Section~\ref{subsec:decision_matrix} maps deployment constraints to the appropriate estimator. Both estimators also run end-to-end on the near-stationary SAPT real-beam acquisitions, in cross-estimator agreement.

These conclusions carry the scope of the present study. The accuracy, robustness, and complementarity results rest on synthetic Schottky spectra with known ground truth; the near-stationary SAPT real-beam acquisitions of Section~\ref{subsec:real_beam} do not exercise the dynamic-tune, deep-low-SNR, or narrow-band regimes in which the two estimators are claimed to differ, so they support feasibility and cross-estimator consistency rather than the regime-dependent comparison itself. The study is also confined to a single facility: the SNR range, revolution-frequency envelope, frame cadence, and operating tune are those of SAPT, and the deployment-fixed constants of both estimators are derived for that regime, so transfer to a machine with substantially different timing or signal quality would require re-deriving them. The reported latency is that of the downstream estimator alone and excludes the shared upstream preprocessing.

The principal future work is operational. Once the SAPT Schottky-monitor prototype's full acquisition chain is developed, commissioned, and brought into routine operation, the two estimators can be evaluated online across the complete injection, acceleration, and extraction cycle, characterising their real-time performance, accuracy, and robustness under actual operating conditions and extending the present feasibility check into the dynamic-tune, deep-low-SNR, and narrow-band regimes that distinguish them.

\bibliographystyle{JHEP}
\bibliography{main}

\end{document}